\documentclass[aps,floats,floatfix,prd,showpacs,twocolumn,nofootinbib]{revtex4}

\usepackage{epsfig}
\epsfclipon
\def\beqa{\begin{eqnarray}}
\def\eeqa{\end{eqnarray}}
\def\beq{\begin{equation}}
\def\eeq{\end{equation}}
\def\subp{+}
\def\subm{-}
\def\subo{{(1)}}
\def\subt{{(2)}}
\def\sgn{{\rm sgn}}
\def\lsim{\mbox{\raisebox{-.6ex}{~$\stackrel{<}{\sim}$~}}}
\def\gsim{\mbox{\raisebox{-.6ex}{~$\stackrel{>}{\sim}$~}}}
\def\sss{\scriptscriptstyle}
\def\alphp{\alpha'}

\begin{document}
\leftline{McGILL-04-06}

\title{Creating the Universe from Brane-Antibrane Annihilation}

\author{Neil  Barnaby, James M.\ Cline}
\affiliation{McGill University, 3600 University St.\
Montr\'eal, Qu\'ebec H3A 2T8, Canada \\
E-mail: barnaby@physics.mcgill.ca, jcline@physics.mcgill.ca}

\date{22 March 2004}

\begin{abstract} When $p$-dimensional branes annihilate with antibranes in the early universe, as
in brane-antibrane inflation, stable $(p-2)$-dimensional branes can appear in the final state.  
We reexamine the possibility that one of these $(p-2)$-branes could be our universe.   In  the
low energy effective theory, the final state branes are cosmic string defects of the  complex
tachyon field which describes the instability of the initial state.   We quantify the dynamics of
formation of these vortices.  This information is then used to estimate the production of
massless gauge bosons on the final branes, due to their coupling to the time-dependent tachyon
background, which would provide a mechanism for reheating after inflation.  We improve upon previous
estimates indicating that this can be an efficient reheating mechanism for observers on the
brane.

\end{abstract}


\maketitle

\section{Introduction}

In the last few years significant progress has been made in constructing string theoretic  cosmological
models where inflation is driven by the naturally occuring potentials between D-branes and their
antibranes \cite{DbraneInflation}-\cite{newinfrefs}. The formation of lower-dimensional branes at the end
of inflation can lead to interesting signals: cosmic-string-like or higher dimensional defects could be
observable remnants  \cite{KibbleMechanism}, possibly providing a rare clue to the stringy origin of
inflation.  A more radical idea was explored in \cite{Cline:Reheating}: perhaps our own observable universe
is such a defect in the higher-dimensional spacetime predicted by string theory.  Since the stable branes
in Type IIB string theory have spatial dimensionalities which are odd, a 3-brane would have descended from
annihilation of 5-branes in this picture, and thus they would be codimension-two defects in of  the
effective 6D theory.  Codimension-two braneworlds have attracted interest lately because of their novel
features, which might have some bearing on the cosmological constant problem \cite{codimension2,CDGV}.

Our interest in this scenario is motivated by questions about the efficiency of reheating in brane-antibrane
inflation \cite{ProblemsWithReheating}.  It is possible that the energy  density liberated from the brane
collisions will be converted mostly into closed string states,  ultimately gravitons, and not necessarily into
visible radiation \cite{ClosedStringRadiation}.  A generic mechanism which could avoid this problem was proposed
in \cite{Cline:Reheating}, wherein the reheating in D-brane driven inflation is due to the coupling of massless
gauge fields to a time-dependent tachyon condensate, which describes the annihilation process.  However, ref.\
\cite{Cline:Reheating} considered only the formation of a tachyon kink instead of the more realistic case of a
vortex, and it used a somewhat crude ansatz for the background tachyon field.   The problem of finding the actual
tachyon background predicted by string theory was studied numerically in  \cite{Cline:DbraneCondensation} but no
attempt was made to improve on the reheating computation. In this paper we aim to analytically determine the
dynamics of formation of lower dimensional branes described as tachyon defects---both kinks and vortices---and to
improve on the reheating calculation of \cite{Cline:Reheating}.

Let us begin by describing the scenario we have in mind.  In the simplest version of D-brane inflation  a 
parallel brane and antibrane begin with some separation between them in one of the extra dimensions.  
Although parallel branes are supersymmetric and have no force between them, the 
brane-antibrane system breaks supersymmetry so that there is an attractive force and hence a nonvanishing 
potential energy.  It is the latter which drives inflation.  Once the branes have reached a critical 
separation one of the stretched string modes between the branes, $T$, becomes tachyonic and the branes
become unstable to annihilation.  The tachyon field starts from the unstable maximum $T=0$ and rolls towards
the vacuum $T \rightarrow \pm \infty$.  However, topological defects may form through the Kibble mechanism 
\cite{KibbleMechanism} so that $T=0$ stays fixed at the core of the defect.  These defects are known to be 
consistent descriptions of branes whose dimension is lower than that of the original branes 
\cite{Sen:TachyonKinkAndVortex,DescentRelations}.  For example, the brane-antibrane system has a complex
tachyon field, leading to vortices which represent codimension-two branes.  On the other hand, an unstable 
brane has a real tachyon which leads to kinks representing codimension-one branes.

The formation of tachyon defects at the endpoint of D-brane inflation is a dynamical process where the
tachyon couples to gauge fields which will be localised on the descendant brane.  It is thus expected that 
some radiation will be produced by the rolling of the tachyon and the problem of reheating becomes 
quantitative: can this effect be efficient enough to strongly deplete the energy density of the tachyon 
fluid so the the universe starts out being dominated by radiation rather than cold dark matter?  It is 
important to stress that though the situation is somewhat analogous to that of hybrid inflation  (where the
tachyon plays the role of the unstable direction in field space which allows for inflation to end  quickly)
the mechanism for reheating is qualitatively different.  The difference is that in the low  energy
effective field theory which describes the tachyon $T$, the potential is minimized at $T=\pm\infty$ and
there are no oscillations about the minimum of the potential. In a normal hybrid inflation model, $T$ would
have a minimum at some finite value and the oscillations of $T$ around its minimum would give rise to
reheating in the usual way.  In the present case, the time dependence of the background is monotonic, not
oscillatory.  Reheating thus might seem to resemble gravitational particle production \cite{quintinf}
rather than the standard picture in which the inflaton decays.  However, in this work we highlight an
important difference between reheating through tachyon condensation and gravitational particle production,
which can make the former much more efficient: there is a divergence in the stress-energy tensor of the
tachyon field within a finite time, which corresponds to the formation of the lower-dimension D-brane.

In this paper we study analytically the dynamical formation of the tachyon vortex and improve  the
reheating calculation, using a slightly simplified model of particle production by tachyon condensation, 
which captures the essential physics revealed by the analysis of vortex formation. In section II we review
the  formation of tachyon kinks which describes the condensation of a brane to a brane of codimension-one. 
In  sections III-V we study analytically the formation of a tachyon vortex on the brane-antibrane pair.  
Section III introduces Sen's action for the complex tachyon field describing this situation. Section IV
presents new analytic results for the time-dependent, complex tachyon field representing vortex formation,
both near to and far from the vortex core.  In section V we show that the stress-energy tensor for the
system splits into a localized, singular piece describing the descendant branes, plus  a bulk contribution
that describes the rolling tachyon condensate.  Section VI introduces the effective action for U(1) gauge
bosons which become localized on the final-state 3-brane, in the rolling tachyon background.  This 
provides a model for the visible radiation produced during reheating. In section VII we calculate the
energy density of this produced radiation on the 3-brane, using some reasonable simplifying assumptions. 
Section VIII gives our conclusions, including speculation about how the final brane-antibrane system could
be stabilized.

\section{Dynamical Tachyon Kink Formation on Unstable Dp-branes}
\label{kinkSection}

In this section we review the dynamical formation of a D$(p-1)$-brane through tachyon condensation on an
unstable D$p$-brane, and derive a few new results. The equations of motion in this case are simpler than in
the case of the vortex and we will use the analysis of this section to reinforce our conclusions when we
analyze the vortex since many of the results are quite analogous.

\subsection{Effective Field Theory and Equations of Motion}

We will work with the effective action for the tachyon on an unstable Dp-brane 
\cite{KinkEffectiveAction,Sen:FieldTheory}
\footnotemark\
\begin{equation}
\label{kinkaction1}
  S = -\int V(T) 
  \sqrt{-\det \left| \eta_{MN} + \partial_M T \partial_N T  \right| } 
  \, d^{p+1} x 
\end{equation}
where we have set the gauge fields and transverse scalars to zero.  We use the potential 
$V(T)=\tau_p \exp \left( -{T^2}/{a^2} \right)$ where
$\tau_p$ is the tension of a D$p$-brane and $a=2\sqrt{\pi \alphp}$.  The value of the constant $a$
is chosen so that the potential satisfies the normalization condition
\begin{equation}
\label{kinkNormalization}
  \int^{+\infty}_{-\infty} V(y)\, dy = 2\pi \sqrt{ \alphp } \tau_p = \tau_{p-1} 
\end{equation}
proposed in \cite{Sen:TachyonKinkAndVortex}.  This normalization was used in \cite{Sen:TachyonKinkAndVortex}
to fix the tension of the singular static kink solution of the action (\ref{kinkaction1}) to correspond to 
the tension of a D$(p-1)$-brane.  For a time-dependent kink solution we take $T$ to be a function of 
$x^{\mu}=(t,x)$ so that the action (\ref{kinkaction1}) becomes
\begin{equation}
\label{kinkaction2}
   S = -\int V(T) \sqrt{ 1 + \partial_\mu T \partial^\mu T} \, d^{p+1}x.
\end{equation}
Static solutions of the theory (\ref{kinkaction2}) are well studied in the literature
\cite{Sen:TachyonKinkAndVortex,StaticKinks}.  Inhomogeneous solutions have also been studied
in some detail \cite{Cline:Reheating,Cline:DbraneCondensation,Felder:Caustics,Inhomogeneous}.  The energy 
momentum tensor for (\ref{kinkaction2}) is\footnotetext{The convention for indices is that upper case roman indices $\{M,N\}$ run over the full
space-time coordinates $\{0,1,\cdots,p\}$, greek indices $\{\mu,\nu\}$ run over the defect
coordinates $\{0,1\}$ and ``hatted'' greek indices $\{\hat{\mu},\hat{\nu}\}$ run over the remaining spatial
coordinates $\{2,3,\cdots,p\}$.  We use metric signature diag$(-1,1,1,\cdots)$.}
\begin{equation}
\label{kinkTmunu}
  T_{\mu\nu} = \frac{V(T)}
  {\sqrt{1 + \partial_\rho T \partial^\rho T}}\partial_\mu T \partial_\nu T - \eta_{\mu\nu}V(T)
  \sqrt{1 + \partial_\rho T \partial^\rho T},
\end{equation}
and the Euler-Lagrange equation of motion is
\begin{equation}
\label{kinkEL}
  \partial^\mu \partial_\mu T - 
  \frac{\partial_\mu \partial_\nu T \partial^\mu T \partial^\nu T}
       {1 + \partial^\rho T \partial_\rho T}
  - \frac{V'(T)}{V(T)} = 0
\end{equation}
where $V'(T)=\frac{\partial V(T)}{\partial T}$.  It is worth noting, as in 
\cite{Ishida:RollingDownToDbrane}, that the equation of 
motion (\ref{kinkEL}) is equivalent to conservation of energy $\partial_{\mu} T^{\mu\nu} =0$ for 
nonconstant $T$ since 
$ \partial_{\mu} T^{\mu\nu} = \partial^{\nu} T
\left[ \partial_{\mu}\left(\frac{\partial \mathcal{L}}{\partial (\partial_{\mu}T )}\right) 
- \frac{\partial \mathcal{L}}{\partial T}  \right] $.  It will be useful in the ensuing analysis to
define
\begin{equation}
\label{kinkSigma}
  \Sigma = \frac{V(T)}{ \sqrt{1+\partial_\mu T \partial^\mu T}  }.
\end{equation}

\subsection{Solutions Near the Core of the Defect}

At the core of the kink we expect the field to stay pinned at $T=0$.  Consider initial data 
$T(t=0,x)=T_i(x)$ and $\dot{T}(t=0,x)=\dot{T}_i(x)=0$
\footnote{For the remainder of this section the dot denotes differentiation with respect to time 
while the prime denotes differentiation with respect to the $x$ coordinate.}.
One expects the field to start to roll where $T_i(x) \not= 0$ due to the small displacement from the 
unstable maximum $V'(T)=0$.  At $t=0$ the equation of motion (\ref{kinkEL}) is

\[
\ddot{T}_i(x) (1+T'_i(x)^2) = T''_i(x) + 2a^{-2}T_i(x)( 1 + T'_i(x)^2 ).  
\]
Clearly any point $x_0$ where
$T_i(x_0)=T''_i(x_0)=0$ will be a fixed point where $\ddot{T}(t,x_0)=0=\dot{T}(t,x_0)$ throughout the 
evolution.  We restrict ourselves only to considering intial data such that 
$\sgn(\ddot{T}_i(x)) = \sgn(T_i(x))$ for all $x$ to ensure that the solutions are increasing.

At the site of the kink (which we take to be $x_0=0$) we have $T=0$; hence there should always be some 
neighbourhood of the point $x=0$ where we can take $V'(T) \cong 0$ so that (\ref{kinkEL}) yields
\begin{equation}
  \ddot{T}(1+T'^2)=(1-\dot{T}^2)T'' + 2\dot{T}T'\dot{T}'.
\end{equation}
This has an increasing solution with $T''=0$
\begin{equation}
\label{Tsmall}
  T(x,t)=x \tan \left[ \frac{\omega}{a}(t-t_c) + \frac{\pi}{2} \right].
\end{equation}
Near the site of the kink the slope of the tachyon field diverges as $(t_c-t)^{-1}$ as $t$ approaches
 the critical time 
$t_c$, similar to the solutions in \cite{Cline:DbraneCondensation}.  

The finite-time slope divergence was observed both numerically and analytically in 
\cite{Cline:DbraneCondensation} and leads to the formation of a singularity in the energy
density at $t=t_c$.  This effect was also found in an exact string theoretic calculation in \cite{Sen:TimeEvolution}.
As $t \rightarrow t_c$ we have $\Sigma(t \rightarrow t_c,x \cong 0) \rightarrow 0$.  The case 
$\Sigma=const$ arises as a first integral of the motion in the static case and the limit 
$\Sigma \rightarrow 0$ corresponds to the singular soliton solution of Sen \cite{StaticKinks}.  It is natural
to expect, then, that as $t \rightarrow t_c$, near $x=0$, the tachyon field $T(t=t_c,x=0)$ coincides with the
stable kink solution of Sen and the time evolution in this neighbourhood stops.  We will argue that at this 
point a codimension-one brane has formed.

\subsection{Vacuum Solutions}
\label{kinkVacuum}

Away from the site of the kink the field is expected to roll towards the vacuum 
$T \rightarrow \pm \infty$ so that $V(T) \rightarrow 0$ at late times for $x \not= 0$.
To analytically study the dynamics near the vacuum it is easiest to work in the
Hamiltonian formalism \cite{Yi:StringFluidTachyonMatter,Hamiltonian} since the Lagrangian vanishes in 
the limit $V(T) \rightarrow 0$, whereas
the Hamiltonian remains well-defined.  Defining the momentum conjugate to $T$ as 
$\Pi={\delta S}/{\delta \dot{T}}$ the Hamiltonian is given by 
$\mathcal{H} = \sqrt{ \Pi^2 + V(T)^2 } \sqrt{ 1 + \partial_i T \partial_i T }$. It is useful to rewrite 
Hamilton's equations of motion, $\dot{\Pi}=-\frac{\partial \mathcal{H}}{\partial T}$ and 
$\dot{T}=\frac{\partial \mathcal{H}}{\partial \Pi}$, in a manifestly covariant way as
\begin{equation}
\label{Hamilton1}
  \partial^\mu T \partial_\mu T + 1 = \frac{V(T)^2}{\Sigma^2}
\end{equation}
and
\begin{equation}
\label{Hamilton2}
  \Sigma\, \partial_\mu \lbrack \Sigma \partial^\mu T \rbrack = V(T)V'(T),
\end{equation}
where $\Sigma= \Pi/\dot{T}$ is defined in (\ref{kinkSigma}).
In the limit $V(T) \rightarrow 0$ equations (\ref{Hamilton1}), (\ref{Hamilton2}) yield
\begin{equation}
\label{kinkEikonal}
  \partial^\mu T \partial_\mu T + 1 = 0
\end{equation}
and
\begin{equation}
\label{SigmaConstraint}
  \partial_\mu \lbrack \Sigma \partial^\mu T \rbrack = 0.
\end{equation}
The solutions of (\ref{kinkEikonal}) were found for arbitrary
Cauchy data in \cite{Felder:Caustics} using the method of characteristics.  The generic solutions 
exhibit the formation of caustics where second and
higher order derivatives become singular.  Caustics are known to form in systems with a pressureless
fluid, which is a good description of the tachyon field as it approaches its ground state $T\to\infty$.
It is not known whether the caustics are a genuine prediction of string theory or just an artifact
of the derivative truncation which leads to the Born-Infeld Lagrangian for the tachyon.  

In any case, caustics are not present in the simplest solution of eq.\ (\ref{kinkEikonal}),
$T = \pm t$, which is the asymptotic form for the homogeneously rolling tachyon.
For $\dot{T}^2 = 1$, eq.\  
(\ref{SigmaConstraint}), which is equivalent to  energy conservation,  implies 
that $\Sigma(t,x) = \Sigma(x)$ is an arbitrary function of $x$.
In this regime the energy momentum tensor 
(\ref{kinkTmunu}) is identical to that of pressureless dust $T_{\mu\nu}=\Sigma(x)u_{\mu}u_{\nu}$ where
$u_{\mu}=\partial_{\mu}T$ is interpreted as the local velocity vector and $\Sigma$ is interpreted as a 
Lorentz-invariant matter density \cite{Sen:FieldTheory,Yi:StringFluidTachyonMatter,Hamiltonian}.

\subsection{Stress-Energy Tensor}

We are interested in the behavior of $T_{MN}$ as $t \rightarrow t_c$.  First we consider the neighbourhood
near $x=0$ where $T(t,x) \cong k {x}/({t_c-t})$ as $t \rightarrow t_c$ (see \ref{Tsmall}).  Near
$x=0$, the Hamiltonian is
\begin{eqnarray*}
 T_{00} & \cong &  \frac{\tau_p k}{t_c-t}\exp\left( -\frac{k^2x^2}{a^2(t_c-t)^2} \right) 
\end{eqnarray*}
and 
\beq
                      \lim_{t \rightarrow t_c^-}      T_{00}   = \sqrt{\pi}a \tau_p \delta(x) 
                           = \tau_{p-1} \delta(x)
\eeq
using the normalization (\ref{kinkNormalization}) for the potential. Similarly
\begin{eqnarray*}
 T_{\hat{\mu}\hat{\nu}}  & \rightarrow &  -\tau_{p-1}\, \delta(x)\, \delta_{\hat{\mu}\hat{\nu}}
\end{eqnarray*}
and $T_{11} \rightarrow 0$ as $x\to 0$.

Consider now the late-time behavior of $T_{MN}$ away from the site of the kink.  Using the solutions
of section \ref{kinkVacuum} we find
$T_{00} \rightarrow \Sigma(x)$ while $T_{11}$, $T_{01}$ and $T_{\hat{\mu}\hat{\nu}}$ tend to zero for 
$x \not= 0$.

To summarize, we find that in the limit of condensation the energy momentum tensor is identical to that of a 
D$(p-1)$-brane:
\begin{eqnarray*}
  T_{00} &=&  \tau_{p-1}\,\delta(x) + \Sigma(x) \\
  T_{11} &=& T_{01} =0                               \\
  T_{\hat{\mu}\hat{\nu}} &=& -\tau_{p-1}\,\delta(x)\,\delta_{\hat{\mu}\hat{\nu}}.
\end{eqnarray*}
The extra bulk energy density $\Sigma(x)$ is similar to the result in 
\cite{Ishida:RollingDownToDbrane} and corresponds to what has been dubbed tachyon matter.

\section{Effective Tachyon Field Theory on the Brane-Antibrane Pair}
\label{Field Theory}

We would like to generalize the results of the previous section to study the dynamical formation of a
tachyon vortex and hence a codimension-two brane.  This is the more realistic situation, since the stable
D-branes of a given string theory are those whose dimensions differ by multiples of two.
We will work with an effective action proposed by Sen \cite{Sen:TachyonKinkAndVortex} for the tachyon on a 
brane-antibrane pair.
The field content for this system is a complex tachyon field T, massless gauge fields 
$A_\mu^{\subo}$, $A_\mu^\subt$ and scalar fields $Y^I_\subo$, $Y^I_\subt$ corresponding to the 
transverse fluctuations of the branes.  The index $(i)=(1),(2)$, which we call the brane index, labels
which of the original branes (actually the brane or the antibrane) the field is associated with.
The effective action is:
\begin{eqnarray}
  S = -\int V(T,Y^I_\subo\!\!\!&-&\!\!\!Y^I_\subt) \left( 
  \sqrt{-\det M^\subo}  \right. \nonumber\\
  &+& \left. \sqrt{-\det M^\subt}  \right)
  \, d^{\,p+1}\! x 
\label{action}
\end{eqnarray}
where \footnote{Greek indices $\{\mu,\nu\}$ are now understood to run over the 
coordinates $\{0,1,2\}$ on which the vortex solutions depend, 
and hatted greek indices $\{\hat{\mu},\hat{\nu}\}$  run over the 
spatial coordinates parallel to the vortex $\{2,3,\cdots,p\}$, where $p=6$ for a vortex which
describes a 3-brane.  Upper case roman indices $\{M,N\}$ 
still run over the full space-time coordinates $\{0,1,\cdots,p\}$.  Finally it will be convenient 
later on to refer to lower case roman indices $\{m,n\}$ which run over only the time and radial 
coordinates $\{0,1\}$. }
\begin{eqnarray}
  M^{(i)}_{MN} &=& g_{MN} + \alphp F^{(i)}_{MN} + \partial_M Y^I_{(i)}\partial_N Y^I_{(i)}
                  \nonumber\\
&+& \frac{1}{2}D_M T D_N T^\ast +  \frac{1}{2}D_M T^\ast D_N T,
\label{Meq}
\end{eqnarray}
\begin{equation}
\label{Feq}
  F^{(i)}_{MN} = \partial_M A^{(i)}_N - \partial_N A^{(i)}_M, \hspace{5mm}
  D_M = \partial_M -iA^\subo_M+iA^\subt_M.
\end{equation}
For the remainder of this paper we will ignore the transverse
scalars and choose $V(T,0)=V(T)=\tau_p \exp \left(-{|T|^2}/{a^2} \right)$ where 
$a$ is chosen so that the static singular vortex solutions of the theory 
(\ref{action}) have the correct tension to be interpreted as codimension 2 D-branes according the 
the normalization proposed in \cite{Sen:TachyonKinkAndVortex}.  We will discuss the normalization
of the potential proposed in \cite{Sen:TachyonKinkAndVortex} in more detail when we calculate the energy
momentum tensor for the theory (\ref{action}).

Though the action (\ref{action}) was not derived from first principles it obeys several necessary 
consistency conditions which are discussed in \cite{Sen:TachyonKinkAndVortex}.  There have been various
other proposals for  the tachyon effective action and vortex solutions on the brane-antibrane pair 
\cite{Tye:CodimensionTwo,Vortices,HN}.  See \cite{Epple:TachyonCondensationAtAngles} for a discussion of 
various models including (\ref{action}).

\section{Vortex Solutions on the Brane-Antibrane Pair}
\label{vortex}

To construct vortex solutions we ignore the transverse scalars and take the remaining fields to 
depend only on the polar coordinates $x^\mu=(x^0,x^1,x^2)=(t,r,\theta)$ with metric 
$g_{\mu\nu} dx^\mu dx^\nu = -dt^2 + dr^2 + r^2 d\theta^2$.  Since the vortex solution should have
azimuthal symmetry we make the ansatz: 
\begin{equation}
\label{ansatz}
  T(t,r,\theta)=e^{i\theta}f(t,r), \hspace{5mm} 
                A_\theta^\subo=-A_\theta^\subt = \frac{1}{2} g(t,r)
\end{equation}
with all other components of $A^{(i)}_\mu$ vanishing.  This generalizes the ansatz 
used in \cite{Sen:TachyonKinkAndVortex} to include time-dependence in the fields $f$ and $g$.  For 
(\ref{ansatz}) one has
\[
  D_t T = e^{i \theta} \dot{f}(t,r), \hspace{3mm}
  D_r T = e^{i \theta} f'(t,r), \hspace{3mm} \]
\[  D_\theta T = e^{i \theta} i (1-g(t,r)) f(t,r) 
\]
and
\[
  F^\subo_{t\theta} = \frac{1}{2} \dot{g}(t,r)=-F^\subt_{t\theta}, \hspace{3mm}
  F^\subo_{r\theta} = \frac{1}{2} g'(t,r)=-F^\subt_{r\theta}
\]
where the dot denotes differentiation with respect to time and the prime now denotes differentiation 
with respect to the radial coordinate.  The matrices $M_{MN}^{(i)}$ are
\begin{eqnarray}
  \left[ M_{MN}^\subo \right] &=& \left[  \begin{array}{cccc}
                                  -1+\dot{f}^2 & \dot{f}f' & \alphp {\dot{g}}/{2} & 0 \\
                                     \dot{f}f' & 1+f'^2 &\alphp {g'}/{2} & 0        \\
                                  -\alphp {\dot{g}}/{2} & -\alphp {g'}/{2} & r^2 + (1-g)^2f^2 & 0 \\
                                           0 & 0 & 0 & \delta_{\hat{\mu}\hat{\nu}}
                                \end{array} \right],   \nonumber\\
  \left[ M_{MN}^\subt \right] &=& \left[ M_{MN}^\subo \right]^{T}.\label{MVeq}
\end{eqnarray}
We also have $\det(M^\subo)=\det(M^\subt)$ since $M^\subo_{MN} = M^\subt_{NM}$ and so we omit
the brane index $(i)$ on $\det(M^{(i)})$ in subsequent calculations.

The action for this ansatz simplifies to:
\begin{eqnarray}
  S  &=& -2 \int V(f) \left( \phantom{\frac{1}{4}\!\!\!\!}
(1+\partial_m f \partial^m f )\left[ r^2 + f^2(1-g)^2 \right]
                       \right.     \nonumber\\
&+& \left. \frac{\alphp^2}{4}\partial_m g \partial^m g 
                            - \frac{\alphp^2}{4} \left( \epsilon^{mn}\partial_m f\partial_n g 
                            \right)^2 \right)^{1/2} \, d^{p+1} x
\label{Lagrangian}
\end{eqnarray}
where $x^m=(x^0,x^1)=(t,r)$ and $g_{mn}dx^m dx^n=-dt^2+dr^2$.  For notational convenience we
define the scalar quantity
\begin{equation}
\label{xi}
  \Sigma(t,r) = \frac{V(f)}{\sqrt{-\det(M)}}
\end{equation}
in analogy with (\ref{kinkSigma}).  The equation of motion for the tachyon is
\begin{eqnarray}
   &\partial_m&\!\!\! \left[ \Sigma \left[ r^2 + (1-g)^2 f^2 \right] \partial^m f 
        - \frac{\alphp^2 \Sigma}{4} ( \epsilon^{ab} \partial_a f \partial_b g ) \epsilon^{mn} \partial_n g
           \right]   \nonumber \\  
                &=&\Sigma ( 1 + \partial^m f \partial_m f )(1-g)^2 f 
                     + \frac{V'(f)V(f)}{\Sigma} ,      \label{eom1}                   
\end{eqnarray}
and the nontrivial component of the equation of motion for the gauge field is
\beqa
  \frac{\alphp^2}{4} \partial_m \left[ \Sigma \partial^m g 
                \right.  \!\!\!&-&\!\! \left.\Sigma ( \epsilon^{ab} \partial_a f \partial_b g ) \epsilon^{mn} \partial_n f
                         \right]\nonumber\\
	&=&         \Sigma ( 1 + \partial^m f \partial_m f )f^2(g-1).
\label{eom2}
\eeqa
Alhough these equations are somewhat cumbersome, inspection of (\ref{eom2}) tells us 
that there should exist a solution $g(t,r)$ such that at $g = 1$, the vacuum, we have 
$\partial_m g = 0$.  This is the asymptotic behavior which corresponds to a vortex solution; 
it is already known from \cite{Sen:TachyonKinkAndVortex} that the static solution $g(r)$ 
is a monotonically increasing function which varies between 0 and 1.  Thus we shall only consider solutions
with the asymptotic behavior 
$ \partial_m g(t,r) \rightarrow 0 \hspace{3mm} \mbox{as} \hspace{3mm} g(t,r) \rightarrow 1$.
We will take initial data $g(0,r)=g_i(r)$ such that 
$g_i(0)=0$, $0 \leq g_i(r) \leq 1$ for all $r$ and $g'_i(r)\geq 0$ for all $r$.  In addition
we will focus on initial tachyon profiles $f(0,r)=f_i(r)$ such that $f_i(0)=0$ and $f'_i(r) > 0$ for all $r$.
For these initial conditions the tachyon must start rolling for $r \not= 0$ due to its
displacement from the unstable vacuum $V'(f)=0$.  Since the asymptotic $g_i(r \rightarrow \infty)
\rightarrow 1$ is an exact solution of (\ref{eom2}) and $g(t,r=0)=0$ by construction, we therefore 
expect $g(r,t)$ to increase towards unity for finite $r \not= 0$.

\subsection{Solutions Near the Core of the Defect}
\label{smallr}

\begin{figure}
\centerline{\epsfxsize=2.95in \epsfbox{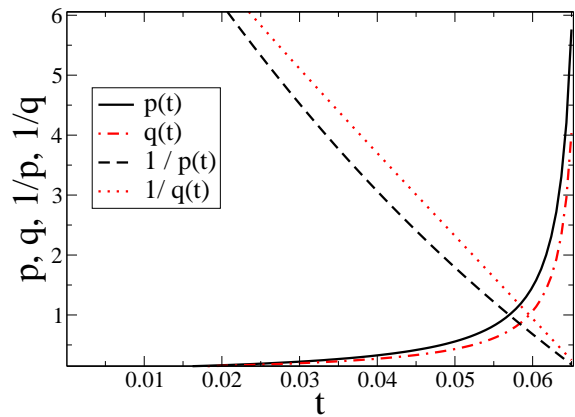}}
\caption{{  Numerical solution for $p(t)$ and $q(t)$ of (\ref{ode1}-\ref{ode2}) 
showing the finite-time slope divergence. $1/p$ and $1/q$  are also shown, demonstrating the
linearity of these functions near the critical
time.
\label{fig1}}}
\end{figure}

As in \cite{Cline:DbraneCondensation}, to analytically study the dynamics near the core of the vortex, 
$r=0$, we make the ansatz:

\begin{equation}
\label{smallfansatz}
  f(t,r) \cong p(t)r, \hspace{5mm}  g(t,r) \cong q(t)r
\end{equation}
for small $r$.  Dropping terms which are subleading in $r$ yields a set of coupled ODEs for $p(t)$ and $q(t)$
\begin{equation}
\label{ode1}
p q^3 \ddot{q} - \ddot{p} q^4 + 2 \dot{p} q^3 \dot{q} - 2 p q^2 \dot{q}^2 + \\
\frac{4}{\alphp^2} p^3 q^2 + \frac{4}{\alphp^2} p q^2 + \frac{2}{a^2} p q^4 =0
\end{equation}
from (\ref{eom1}) and 
\beqa
-p\ddot{p} q^3 \!\!\!&+&\!\! p^2 q^2 \ddot{q} + q^2 \ddot{q} - 2 q \dot{q}^2 + 2 p \dot{p} q^2 \dot{q} 
- 2 p^2 q \dot{q}^2 + \frac{4}{\alphp^2} p^4 q \nonumber\\
&+& \frac{8}{\alphp^2} p^2 q + \frac{4}{\alphp^2} q + \frac{2}{a^2} p^2 q^3 =0
\label{ode2}
\eeqa
from (\ref{eom2}).  Although (\ref{ode1},\ref{ode2}) are difficult to solve analytically it is 
straightforward to verify that in the regime where $\dot{p}$, $\dot{q}$ and higher derivatives
are large compared to $p$, $q$ there exists an approximate solution to (\ref{ode1}-\ref{ode2}) 
where both $p(t)$ and $q(t)$ are divergent in finite time $t_c$:
\begin{equation}
\label{divergences}
  p(t) = \frac{p_0}{t_c-t}, \hspace{5mm} q(t)= \frac{q_0}{t_c-t}
\end{equation}
in analogy with the kink solution of \cite{Cline:DbraneCondensation}.  
Numerical solutions to (\ref{ode1}-\ref{ode2}) agree with this prediction, as shown in figure
\ref{fig1}.

\subsection{Solutions Away From the Core of the Defect}
\label{larger}

We are interested in solutions where $\partial_m g \rightarrow 0$ as 
$g \rightarrow 1$ and where $f(t,r) \rightarrow \infty$ as $t \rightarrow \infty$.  Since, as we have seen 
above, $g'(t,r=0)$ is diverging in finite time, therefore $g$ must be increasing to unity for $r \not= 0$ so
that at late times we expect $g(t,r)$ to resemble a step function.  Thus to study the dynamics away from the
core of the defect we begin with the ansatz:
\begin{equation}
\label{sigma}
  g(t,r)=1-\varepsilon \sigma(t,r)
\end{equation}
and work only to leading order in $\varepsilon$.  The leading order contribution to (\ref{eom1})
decouples completely from $\sigma(t,r)$
\begin{eqnarray}
   \left[ \ddot{f} (1+f'^2)\right. \!\!\!&-&\!\!\! \left. f'' (1-\dot{f}^2) -2\dot{f}f'\dot{f}' 
          -\frac{2f}{a^2}(1-\dot{f}^2+f'^2) \right] r   \nonumber  \\ 
   &-& \left[ 1-\dot{f}^2 + f'^2 \right] f'  = 0.   \label{pde1}
\end{eqnarray}
We can consistently find solutions of (\ref{pde1}) by taking $f$ to be a solution of the eikonal
equation $1-\dot{f}^2 + f'^2=0$.  Subject to this constraint the second term in the square braces
in (\ref{pde1}) vanishes trivially.  The constraint that the first term in the square braces in (\ref{pde1})
vanishes is exactly the same as the equation of motion one
would derive from  $\mathcal{L}=-V(f)\sqrt{1+\partial_m f \partial^m f}$, the 
Born-Infeld Lagrangian.  The eikonal equation yields the Born-Infeld equation as a differential
consequence, which is not surprising since this amounts to minimizing the action by setting 
$\mathcal{L}=0$.  Thus the PDE (\ref{pde1}) is automatically satisfied when $f$ is a solution of the
eikonal equation.  We find, then, that when $g(t,r) \cong 1$ the tachyon field must obey
\begin{equation}
\label{eikonal}
  \partial_m f \partial^m f + 1 = 0
\end{equation}
as in section \ref{kinkSection}.

The ansatz (\ref{sigma}) yields no simplification of (\ref{eom2}); however, we may solve for
$\sigma(t,r)$ given (\ref{eikonal}) based on more fundamental constraints.  The argument of 
the square root in (\ref{Lagrangian}) must be nonnegative to ensure reality of the Lagrangian.
Thus for $f(t,r)$ given by (\ref{eikonal}) the requirement that the Lagrangian be real translates
to
\[
  \partial_m \sigma \partial^m \sigma -( \epsilon^{mn} \partial_m f \partial_n \sigma )^2 \geq 0. 
\]
or, with $f$ given by (\ref{eikonal}),
\begin{equation}
\label{realL}
  - \left( \dot{\sigma} \sqrt{1+f'^2} - \sigma' f' \right)^2  \geq 0.
\end{equation}
For real fields it is clear that (\ref{realL}) can only be solved when the equality is taken.  
It is worth noting that since the Lagrangian vanishes when the equality is taken in (\ref{realL}), this
constraint ensures that the full equations of motion are satisfied.

As in section \ref{kinkSection} we can avoid the difficulties of caustic formation in the general 
solutions of (\ref{eikonal}) found in \cite{Felder:Caustics} by taking the Cauchy data to be linear and 
using the one-parameter family of solutions
\begin{equation}
\label{linearsolution}
  f(t,r)=\alpha t + \sqrt{\alpha^2-1}\, r.
\end{equation}
Reality of the Lagrangian requires
\[
  \alpha \dot{\sigma} - \sqrt{\alpha^2-1}\, \sigma' = 0.
\]
This PDE is separable and we find the solution
\begin{equation}
\label{sigmasolution}
  g(t,r) = 1 - \varepsilon \exp \left( -\frac{r}{R} \right)
                           \exp \left( -\frac{t}{R} \sqrt{ \frac{\alpha^2-1}{\alpha^2}} \right)
\end{equation}
where $R$ is a separation constant.  Note that the solution (\ref{sigmasolution}) becomes static 
in the limit $\alpha^2 \rightarrow 1$, the homogeneous rolling tachyon.  In fact, for $\alpha^2=1$ any 
function $\sigma(t,r)=\sigma(r)$ satisfying the necessary boundary condition $\sigma(r \rightarrow \infty) \rightarrow 0$ 
will generate a solution.  We will ultimately be interested in this limit.\footnote{The exact functional
form of $\sigma$ at large $r$ and late times turns out to be of little importance to the ensuing analysis.}

It is noteworthy that taking $g(t,r)$ close to unity (\ref{sigma}) ultimately translates into the 
requirement that $\det(M)$ must vanish.  The solutions
(\ref{linearsolution}-\ref{sigmasolution}) should be thought of as late-time asymptotics where
$V(f) \rightarrow 0$ since  $f \rightarrow \infty$.  In this limit $\Sigma(t,r)$ defined in
(\ref{xi}) has the indeterminant form  $\frac{0}{0}$ as in section \ref{kinkSection}.  The
quantity $\Sigma$ is of some interest for two reasons.   First, it parametrizes the manner in
which we take the limits $V(f) \rightarrow 0$ and  $\det(A) \rightarrow 0$ as we approach the
vacuum state.  Second, the form of $\Sigma$ for $r \not=0$ will  determine the form of the
energy-momentum tensor in that regime, as in the case of the kink.  Following the discussion in
section \ref{kinkSection} we will use energy-momentum conservation to place constraints
on the aymptotic form of $\Sigma$.

\section{Stress-Energy Tensor}
\label{EnergyMomentumTensorSection}
In this section we demonstrate that the vortex solutions found above give rise to the
formation of a singularity in the stress-energy tensor of the tachyon field, which corresponds
exactly with that of a codimension-two D-brane in the final state, whose tension has the value
expected from string theory.  We also derive the bulk stress-energy tensor for the leftover
tachyon matter, which continues to roll even after the formation of the D-brane.

The stress-energy tensor for the action (\ref{action}) is
\beqa
  T^{MN} &=& -\frac{V(T,Y^I_\subo-Y^I_\subt)}{r} \left[ 
               \sqrt{-\det M^\subo}\big( M^{-1}_\subo\big)^{MN}_S \right. \nonumber\\
 &+&
            \left.   \sqrt{-\det M^\subt} \big( M^{-1}_\subt \big)^{MN}_S
               \right]
\eeqa
where the subscript $S$ denotes the symmetric part of the matrix, {\it i.e.,} 
$( M^{-1}_{(i)})^{MN}_S = \frac{1}{2} [ 
( M^{-1}_{(i)})^{MN} + ( M^{-1}_{(i)} )^{MN} ]$.
The components of $T^{MN}$ parallel to the vortex simplify to
\begin{equation}
\label{transverse}
  T^{\hat{\mu}\hat{\nu}} = -\frac{2}{r}V(f) \sqrt{-\det(M)}\delta^{\hat{\mu}\hat{\nu}}.
\end{equation}
For the components involving $t,r,\theta$, it is 
useful to rewrite $T^{\mu\nu}$ in terms of $\Sigma$ and the symmetrized cofactor 
matrix of $M_{\mu\nu}^{(i)}$, which we define as $C^{\mu\nu}_{(i)}$.\footnote{That is to say $C^{\mu\nu}_{(i)} = \det(M) (M^{-1}_{(i)})^{\mu\nu}_{S}$.}\ Since 
$C^{\mu\nu}_\subo=C^{\mu\nu}_\subt$ we drop the brane index on the 
cofactor matrices.  In the $t,r,\theta$ directions, then, we have
\begin{equation}
\label{energy-momentum}
  T^{\mu\nu} = \frac{2\Sigma}{r}  C^{\mu\nu}.
\end{equation}
The nonzero components of the cofactor matrix for the Lagrangian (\ref{Lagrangian}) are
\begin{eqnarray*}
  C^{tt} &=& \left[ r^2 + f^2 (1-g)^2 \right](1+f'^2) + \frac{\alphp^2}{4}g'^2,  \\
  C^{tr} &=& -\left[ r^2 + f^2 (1-g)^2 \right]\dot{f}f' - \frac{\alphp^2}{4}\dot{g}g',  \\
  C^{rr} &=& -\left[ r^2 + f^2 (1-g)^2 \right](1-\dot{f}^2) + \frac{\alphp^2}{4}\dot{g}^2,  \\
  C^{\theta\theta} &=& -( 1 - \dot{f}^2  + f'^2 ).
\end{eqnarray*}

\subsection{Normalization of the Potential}

In \cite{Sen:TachyonKinkAndVortex} Sen finds that the action (\ref{action}) provides a good effective 
description of the 
tachyon on the brane-antibrane system provided the potential $V(T)$ is chosen to satisfy the
normalization constraint 
\begin{equation}
\label{normalization}
  \tau_{p-2} = 4 \pi \int_0^\infty V(z) \sqrt{ z^2(1-\hat{G}(z))^2+\frac{\alphp^2}{4}(\hat{G}'(z))^2 } 
                     \, dz 
\end{equation}
where $\tau_{p-2}= (2 \pi)^2 \alphp \tau_p$ is the tension of a ($p-2$)-brane, $\hat{G}(z)=G(F^{-1}(z))$,
and $\{f(r)=F(br),\ g(r)=G(br)\}$ are the static soliton solutions to be understood in the limit that 
$b \rightarrow \infty$.  This constraint is necessary to
ensure that the vortex solution has the correct tension to be interpreted as a D$(p-2)$-brane.  In the 
time-dependent case there is some ambiguity as to how to interpret (\ref{normalization}) since 
this statement appears to depend on the functional form of the solutions, which would make the right-hand-side
apparently time-dependent.\footnote{Arguments are presented in \cite{Sen:TachyonKinkAndVortex} for why (\ref{normalization}) is
in fact only a constraint on $V(T)$ and not on the solutions $T(x)$ themselves.}\

But physically, it makes sense to impose (\ref{normalization}) at 
$t\ge t_c$ since $t_c$  is the time by which the brane has actually formed, and in the limit 
$t \rightarrow t_c$ the time-dependent solutions should coincide with the soliton solutions 
in the neighbourhood of $r=0$.  In the time-dependent case, it is $(t_c-t)^{-1}$ which tends to  
infinity (as $t \rightarrow t_c$)
and plays the role of $b$.  Although the exact functional forms of $G(z)$, $F(z)$ are not known for all 
$z$, we can infer from (\ref{divergences}) that $G(z) \cong q_0 z$ and $F(z) \cong p_0 z$ near $z=0$.
Furthermore, we know that $G(z) \rightarrow 1$ for sufficiently large $z$ by construction.  
We also have $F(z=0)=0$ and $F(z \not= 0) \not= 0$ so that $F^{-1}(0)=0$ and $F^{-1}(z \not= 0) \not= 0$.

Let us consider the two terms under the square root in (\ref{normalization}). The first term, 
$z^2(1-\hat{G}(z))^2$, is small near $z=0$ due to the overall multiplicative factor of $z^2$.  On the other
hand, at large $z$ this term is also small since $F^{-1}(z)$ is large and hence 
$\hat{G}(z) = G(F^{-1}(z)) \cong 1$.  We conclude that the derivative term under the root in 
(\ref{normalization}) dominates.  At small $z$ we have $F^{-1}(z) \cong p_0^{-1} z$ and thus
$\hat{G}(z) \cong q_0 F^{-1} (z) \cong \frac{q_0}{p_0} z$ so that $\hat{G}'(z) \cong \frac{q_0}{p_0}$.
For simplicity we take $\hat{G}'(z) \cong \frac{q_0}{p_0}$ for all $z$ since this expression 
is multiplied by $V(z)$ which tends to zero quickly for large $z$.  We find then that
\begin{equation}
\label{normalization2}
    \frac{q_0}{p_0} \int_{-\infty}^\infty e^{-{z^2}/{a^2}} \, dz = 4 \pi.
\end{equation}
The normalization (\ref{normalization2}) is 
equivalent to $a = 4 \sqrt{\pi} \frac{p_0}{q_0}$.  We shall see later on that the relation 
$a = 4 \sqrt{\pi} \frac{p_0}{q_0}$ may be equivalently viewed as a constraint on the arbitrary
function $\Sigma(t,r)$.

\subsection{Stress-Energy Tensor at $r=0$}

At $r \cong 0$ and $t \rightarrow t_c$ the solutions (\ref{divergences}) are valid and the 
Hamiltonian is
\begin{eqnarray*}
  T^{00} &=& \frac{2 \Sigma}{r} \left( \left[ r^2 + f^2 (1-g)^2 \right](1+f'^2) 
             + \frac{\alphp^2}{4}g'^2 \right) \\
         &\cong& \frac{\tau_p}{r} \frac{q_0 \alphp}{t_c-t} 
                   \exp \left( -\frac{p_0^2 r^2}{a^2(t_c-t)^2} \right)
\end{eqnarray*} {to leading order in $r$}. Then,  using $a=4 \sqrt{\pi}\frac{p_0}{q_0}$,
\begin{eqnarray*}
      \lim_{t \rightarrow t_c} T^{00}   &=& 4 \pi \tau_{p-2} \, \frac{\delta(r)}{r} \hspace{5mm}\\
         &=& \tau_{p-2} \, \delta(r \cos\theta)\, \delta( r \sin\theta )
\end{eqnarray*}
and the components parallel to the vortex are
\begin{eqnarray*}
  T^{\hat{\mu}\hat{\nu}} &=& -\frac{2}{r} \delta^{\hat{\mu}\hat{\nu}} V(f) \sqrt{-\det(M)}  \\
                         &\rightarrow& -\tau_{p-2}\, \delta^{\hat{\mu}\hat{\nu}}
                                       \delta(r \cos\theta)\, \delta( r \sin\theta ).
\end{eqnarray*}
The remaining components, $T^{11}$ and $T^{01}$, are vanishing at $r=0$.
The angular component $T^{22}=T^{\theta\theta}$ contains a delta function at $r=0$; however this is an
artifact of $\theta$ being a bad coordinate at $r=0$; it can be seen that $T_{\theta\theta}=0$, and going
to Cartesian coordinates confirms that the $T_{\mu\nu}=0$ for the transverse coordinates, as should be the
case for a D-brane. This result is the same as in the static case \cite{Sen:TachyonKinkAndVortex}.

\subsection{Stress-Energy Tensor at $r>0$}

For $r>0$ at late times the solutions (\ref{linearsolution},\ref{sigmasolution}) are valid.  
For simplicity we take 
$\alpha^2=1$ and work only to leading order in $\varepsilon$.
The Hamiltonian is
\begin{eqnarray*}
  T^{00}  &\cong& 2r\Sigma(t,r)
\end{eqnarray*}
and the remaining components of $T^{MN}$ vanish at late times for $r>0$.
Notice that conservation of energy $\partial_M T^{MN} = 0$ at large $r$ forces 
\[
\Sigma(t,r) = \Sigma(r).
\]
That is, $\Sigma$ is an arbitrary function of $r$ as in section \ref{kinkSection}.

\vspace{3mm}

To summarize, we find that in the limit of condensation the energy momentum tensor is identical
to that of a D$(p-2)$-brane
\begin{eqnarray}
  T^{00} &=& \tau_{p-2}\,\delta(r\cos\theta)\,\delta(r\sin\theta) + 2r\Sigma(r)  \nonumber \\
  T^{11} &=& T^{22} = 0 \label{finalTmunu} \\
  T^{\hat{\mu}\hat{\nu}} &=&  -\delta^{\hat{\mu}\hat{\nu}} 
                               \tau_{p-2}\delta(r\cos\theta)\,\delta(r\sin\theta) \nonumber
\end{eqnarray}
with all off-diagonal components vanishing. The extra bulk energy density $2r\Sigma(r)$ is similar to 
the result in \cite{Ishida:RollingDownToDbrane} and section \ref{kinkSection} and corresponds to 
tachyon matter rolling toward $T\to\infty$ in the bulk.

\subsection{Conservation of Energy}

We can constrain $\Sigma(r)$ using conservation of energy.  Initially the system 
consists of two D$p$-branes with energy density $2 \tau_p V_2$  in the ($p-2$)-dimensional subspace spanned
 by
$\{x^{\hat{\mu}}\}$, where $V_2$ is the volume of the 2-dimensional subspace spanned
by $\{r,\theta \}$.  At  late times, after the codimension 2 brane and its antibrane
have formed, the energy density in the
($p-2$)-dimensional space is given by the sum of the D($p-2$)-brane tensions, $2\tau_{p-2}$,  and the energy
density due to  tachyon matter $4 \pi \int r^2 \Sigma(r) \, dr$.  Conservation of energy thus implies
\begin{eqnarray*}
  2 \tau_p V_2 &=& 2\tau_{p-2} + 4 \pi \int dr \, r^2 \Sigma(r) \\
  \left( 2 V_2 - 2 (2 \pi)^2 \alphp \right) \tau_p &=& 4 \pi \int dr \, r^2 \Sigma(r).
\end{eqnarray*}
Since $\Sigma(r)$ is arbitrary we can take 
\begin{equation}
\label{Xisplit}
  \Sigma(r) = \frac{\tau_p}{r} + \frac{\tilde{\Sigma}(r)}{4 \pi r^2}
\end{equation}
where $\tilde{\Sigma}(r)$ satisfies the constraint
\begin{equation}
\label{Xitwiddle}
  \int dr \, \tilde{\Sigma} (r) = -8\pi^2 \alphp \tau_p.
\end{equation}
The conditions (\ref{Xisplit}),(\ref{Xitwiddle}) are equivalent to (\ref{normalization2}) and may be 
thought of as an alternative to the normalization (\ref{normalization}).

\section{Inclusion of Massless Gauge Fields}

We will now restrict ourselves to a (5+1)-dimensional spacetime with $\{M,N\}$=$\{0,1,\cdots,5\}$,
$\{\mu,\nu\}$=$\{0,1,2\}$ and $\{\hat{\mu},\hat{\nu}\}$=$\{3,4,5\}$.  There are two gauge fields in the
problem: $A^{M}_\subo$ and $A^{M}_\subt$, or equivalently $A^{M}_\subp=A^{M}_\subo+A^{M}_\subt$ and
$A^{M}_\subm=A^{M}_\subo-A^{M}_\subt$, which have different couplings to the tachyon.  We have already
shown that $A^{\mu}_\subm$ is the field which condenses in the vortex, hence its associated gauge symmetry is
spontaneously broken. For reheating it is thus $A^{\hat{\mu}}_\subp$ which most closely resembles the
Standard Model photon.  We will ignore fluctuations of the  heavy fields $A^{\hat{\mu}}_\subm,$
$A^{\mu}_\subp$, and $A^{\mu}_\subm$, keeping only the background solution for $A^{\mu}_\subm$ (which was
given in section \ref{vortex}), and the fluctuations of the photon  $A^{\hat{\mu}}_\subp$.  This leads to
considerable simplification since it ensures  that $D_{\hat{\mu}}T=0$.  To compute the production of 
photons in the time-dependent background, we want to expand the action
(\ref{action}) to quadratic order in $A^{\hat{\mu}}_\subp$.

The matrix $M_{MN}^{(i)}$ of eq.\ (\ref{Meq}) can be written in block diagonal form as
\begin{equation}
\label{matrix}
    \left[ M_{MN}^{(i)} \right] = \left[  \begin{array}{cc}
        \left[ \mathcal{V}_{\mu\nu}^{(i)} \right] & \left[ S_{\mu\hat{\nu}}^{(i)} \right]   \\
-\left[ S_{\mu\hat{\nu}}^{(i)} \right]^T & 
        \left[ \delta_{\hat{\mu}\hat{\nu}} + \alphp F_{\hat{\mu}\hat{\nu}}^{(i)} 
           \right] \\
                                           \end{array} \right]   \\
\end{equation}
where $\mathcal{V}_{\mu\nu}^{(i)}=M_{\mu\nu}^{(i)}$  is the 
contribution from the vortex background given in (\ref{MVeq}),
$S_{\mu\hat{\nu}}^{(i)}= \alphp \partial_\mu A_{\hat{\nu}}^{(i)}$ is the contribution from 
$\{t,r,\theta\}$ derivatives of $A_{\hat{\mu}}^{(i)}$, and $F_{\hat{\mu}\hat{\nu}}^{(i)}$
is the field strength tensor for $A_{\hat{\mu}}^{(i)}$.  Using a well-known identity for 
determinants we can write 
\begin{equation}
\label{determinantidentity}
\det(M^{(i)})=\det\mathcal{V}^{(i)}\det\left( {\bf 1}+ \alphp F^{(i)} + 
                           S_{(i)}^T \mathcal{V}^{-1}_{(i)} S_{(i)} \right).
\end{equation}
Expanding $\det( {\bf 1} + \alphp F^{(i)} +  S_{(i)}^T \mathcal{V}^{-1}_{(i)} S_{(i)} )$ to quadratic 
order in $A_{\hat{\mu}}^{(i)}$, the action (\ref{action}) becomes
\begin{eqnarray}
  S  &\cong& - \alphp^2 \frac14\int \, d^{p+1} x  V(f)\sqrt{-\det \mathcal{V} } \left( 
              (\mathcal{V}^{-1})^{\mu\nu}_{(S)} \partial_\mu A_{\hat{\mu}}^\subp 
                                                      \partial_\nu A^{\hat{\mu}}_\subp\right.
	\nonumber\\
 &+& \left.\partial_{\hat{\mu}} A_{\hat{\nu}}^\subp 
\partial^{\hat{\mu}} A^{\hat{\nu}}_\subp\right) 
         \label{transverseLagrangian}
\end{eqnarray}
where we have chosen the gauge $\partial_{\hat{\mu}} A^{\hat{\mu}}_\subp=0$ and disregarded the piece which
does not depend on $A^{\hat{\mu}}_\subp$.  We omit the brane index on $\mathcal{V}_{(i)}$ since the 
determinant and the symmetric part of $\mathcal{V}^{-1}_{(i)}$ are equal for both $i=\{1,2\}$.
Defining an effective metric $G_{MN}$ by
\begin{equation}
\label{effectivemetric}
    \left[ G_{MN} \right] = \left[  \begin{array}{cc}
        \left[ \mathcal{V}_{\mu\nu}^{(S)} \right] & \left[ 0 \right]   \\
\left[ 0 \right] & 
        \left[ \delta_{\hat{\mu}\hat{\nu}}  \right] \\
                                           \end{array} \right]   \\
\end{equation}
the action (\ref{transverseLagrangian}) may be written as
\begin{equation}
\label{effectiveaction}
   S=-\frac{\alphp^2}{4} \int V(f)  
    \sqrt{-G} G^{MN}\delta^{\hat{\mu}\hat{\nu}}\partial_M A_{\hat{\mu}}^\subp \partial_N A_{\hat{\nu}}^\subp
     \, d^{p+1} x. 
\end{equation}
From (\ref{effectiveaction}) one sees that the fluctuations of the photon behave like a collection of
massless scalar fields propagating in a nonflat spacetime described by the metric $G_{MN}$, with a
position- and time-dependent gauge coupling given by  $g^2 = 1/V(f(t,r))$.

To get an intuitive sense for the behavior of the action (\ref{effectiveaction}) we note that the 
stress-energy  tensor derived in section \ref{EnergyMomentumTensorSection} can be written as
\[
  T^{MN} = - \frac{2}{r} V(f) \sqrt{-G} G^{MN}.
\]
In the limit of condensation, $T^{MN}$ is given by (\ref{finalTmunu}), so that once the brane has formed
the action (\ref{effectiveaction}) reduces to a description of gauge fields propagating in a 
(3+1)-dimensional Minkowski space, with an additional component which couples the gauge fields to the 
tachyon matter density in the bulk.  In other words, the effective metric $G_{MN}$ starts off being
smooth throughout the bulk, but within the time $t_c$, its support collapses to become a delta function
$\delta^{(2)}(\vec x)$ in the relevant extra dimensions $\{r,\theta\}$.

The equations of motion resulting from the effective action (\ref{effectiveaction}) are difficult to solve
analytically since the effective metric $G_{MN}$ depends nontrivially on both $r$ and $t$ and is 
nondiagonal in the subspace of $\{t,r\}$.  For this reason we would like to propose a simplified model of
the condensation which captures the essential features of the action (\ref{effectiveaction}).  We have 
derived solutions for the vortex background valid at small $r$, $r \lsim (t_c-t)$,
and at large $r$, $r \gsim (t_c-t)$.  
Similarly, the energy momentum tensor we have derived corresponding to these 
solutions has very different behavior in the $r \leq (t_c-t)$ and $r > (t_c-t)$ regions of the spacetime.  

For $r \leq (t_c-t)$ the energy momentum tensor contracts to a delta function centered at 
$r=0$, with $(t_c-t)$ playing the role of the small parameter which regularizes the delta function.  
That is to say, 
\[
r T^{MN} \cong \frac{1}{t_c-t} \exp \left(-\frac{p_0^2r^2}{a^2(t_c-t)^2} \right) H^{MN}
\]
at small $r$, where the matrix entries $H^{00}$, $H^{\hat{\mu}\hat{\nu}}$ are finite as $t \rightarrow t_c$
and the remaining components of $H^{MN}$ tend to zero (near $r=0$) as $t \rightarrow t_c$.

In the $r > (t_c-t)$ region, the energy momentum tensor has quite different behavior. 
After condensation of the defect has completed at $t=t_c$, the
energy density in the bulk ($r>0$) is due entirely to tachyon matter, 
while the part of the stress-energy which is going into the tension of the defect
vanishes in this region.  Hence
\[
r T^{MN} \rightarrow 2 \delta^M_0 \delta^N_0 r^2 \Sigma(r)
\]
as $t \rightarrow t_c$, for $r > (t_c-t)$.

In our simplified model of the particle production due to the tachyon condensation we
therefore split the energy momentum tensor into brane and bulk pieces $T^{\sss MN}  =
T^{\sss MN}_{\rm brane} + T^{\sss MN}_{\rm bulk}$ where $T^{\sss MN}_{\rm brane}$
contracts to a delta function as $t \rightarrow t_c$ and  $T^{\sss MN}_{\rm bulk}
\rightarrow  2 \delta^{\sss M}_0 \delta^{\sss N}_0 r \Sigma(r)$ in the same limit.  The action 
(\ref{effectiveaction}) then splits into two components $S = S_{\rm bulk} + S_{\rm brane}$. 
We expect most of the particle production to occur near the end of the
condensation, when the background tachyon field is becoming singular near the vortex,
so the best approximations for the simplified gauge field action
 are those which describe the exact expression most accurately near $t=t_c$:
\begin{eqnarray*}
 \!\!\!\!\!\!\!\!\!\!\!\!  S_{brane} &\propto&   - \int  \frac{1}{t_c-t} 
	\exp \left(-\frac{p_0^2r^2}{a^2(t_c-t)^2} \right) \\
        &&\phantom{.}\qquad H^{MN} \delta^{\hat{\mu}\hat{\nu}} \partial_{M} A^{\subp}_{\hat{\mu}} 
       \partial_{N} A^{\subp}_{\hat{\nu}} \, dt dr d\theta dx^{\hat{\alpha}}
\end{eqnarray*}
and
\[
  S_{bulk} \propto -\int r^2 \Sigma(r) 
          \left( -\delta^{\hat{\mu}\hat{\nu}} \dot{A}^{\subp}_{\hat{\mu}} \dot{A}^{\subp}_{\hat{\nu}} \right)
          \, dt dr d\theta dx^{\hat{\alpha}}.
\]
At earlier times the coefficient of the bulk part of the Lagrangian 
would have time dependence,  and the bulk Lagrangian would
contain contributions from all the derivative of the gauge field,
but this form is valid close to $t_c$.
 
Finally we argue that the bulk part of the action can be ignored.  To this end,
let us change to coordinates which are comoving
with the contraction of the vortex core:
\beq
\label{rtilde}
  \tilde{r}=\frac{r}{t_c-t}, \hspace{5mm} \tilde{t} = t_c-t.
\eeq
In terms of these coordinates the ``small $r$'' solutions are valid for $\tilde{r} \leq 1$ and the 
``large $r$'' solutions are valid for $\tilde{r}>1$.  The Jacobian of this transformation is $-\tilde{t}$
so that
\beqa
&&  S_{\rm brane} \propto \nonumber\\
&&-\int  \exp \left(-\frac{p_0^2\tilde{r}^2}{a^2} \right) 
         H^{MN} \delta^{\hat{\mu}\hat{\nu}} \partial_{M} A^{\subp}_{\hat{\mu}} 
         \partial_{N} A^{\subp}_{\hat{\nu}} \, 
                    d\tilde{t} d\tilde{r} d\theta dx^{\hat{\alpha}}.\nonumber
\eeqa
To lowest order in $\tilde{r}^2$ the matrix entries $H^{MN}$ in terms of these new coordinates are all 
constant.  Consider now the piece of the action which couples to the tachyon matter 
density in the bulk, written in terms of these new coordinates:
\[
 S_{\rm bulk} \propto -\int \tilde{r}^2 \tilde{t}^2 \Sigma(\tilde{r}\tilde{t}) 
          \left( -\delta^{\hat{\mu}\hat{\nu}} \dot{A}^{\subp}_{\hat{\mu}} \dot{A}^{\subp}_{\hat{\nu}} \right)
          \, d\tilde{t} d\tilde{r} d\theta dx^{\hat{\alpha}}.
\]
Since $\Sigma(z)$ is not singular at $z=0$, the bulk piece of the action become
negligible near the end of the contraction $\tilde{t} \rightarrow 0$.  This is a
consequence of the fact that as the condensation  proceeds the gauge field is confined to
the descendant brane.

\section{Simplified Model of Reheating}

In the previous section we argued that the gauge field couples most strongly to the part
of the tachyon background which is collapsing to form the defect.  This closely resembles
a gauge theory defined on a manifold in which a two-dimensional subspace which is
shrinking with time. As a simplified model of the interaction we thus consider a massless
spin-1 field 
\[
S=-\frac{1}{4}\int  
    \sqrt{-g} g^{MA} g^{NB} F_{MN} F_{AB}
     \, d^{p+1} x
\]
propagating in a FRW-like background
\begin{equation}
\label{toymetric}
  g_{MN}dx^M dx^N = -dt^2 + R(t)^2\left(d\tilde r^2+\tilde r^2d\theta^2 \right) + 
                     \delta_{\hat{\mu}\hat{\nu}}dx^{\hat{\mu}} dx^{\hat{\nu}}.
\end{equation}
The coordinate $\tilde r$ in (\ref{toymetric}) is fixed with the expansion and  is thus
corresponds to $\tilde{r}$ defined in (\ref{rtilde}).   However for simplicity of
notation we will drop the tilde and write $r$ instead of $\tilde r$ in the remainder of
the paper.  We take $r$ to be dimensionless while $t$, $x^{\hat{\mu}}$ and $R$
have  dimensions of length.  

If we restrict ourselves to configurations with $A^{\mu}=0$, $A^{\hat{\mu}} \not= 
0$\footnote{This restriction will only underestimate the reheating.  Since we want to
show that the reheating can be efficient, this approximation will not weaken
the ensuing argument.} and impose the gauge condition $\partial_{\hat{\mu}} A^{\hat{\mu}}
= 0$  then the action simplifies to 
\begin{equation}
\label{toyaction}
S=-\frac{1}{2}\int  
    \sqrt{-g} g^{MN}\delta^{\hat{\mu}\hat{\nu}}\partial_M A_{\hat{\mu}} \partial_N A_{\hat{\nu}}
     \, d^{p+1} x
\end{equation}
Notice that for the metric (\ref{toymetric}) and the gauge field configuration 
$A^{\mu}=0$ we have 
chosen, one has  $\nabla_{M} A_{\hat{\mu}} = \partial_{M} A_{\hat{\mu}}$ and
$R_{MN} A^M A^N = R_{\hat{\mu}\hat{\nu}} A^{\hat{\mu}} A^{\hat{\nu}} =0$.

We will impose homogeneous boundary conditions at $r=1$ and take the scale factor in (\ref{toymetric}) to be
\beq
\label{Req}
  R(t) = \left \{ \begin{array}{ll} R_0 & \mbox{if $t < 0$}; \\
          R_0-\eta t   & \mbox{if $0 \leq t \leq t_c$}; \\ 
          R_0-\eta t_c = \epsilon  & \mbox{if $t>t_c$}.  \end{array} \right.
\eeq
where $R_0$ represents the initial radial size of the extra dimensions. This approximation for the
time-dependence of the vortex core is the simplest  form which has the same qualitative behavior as
the true background, while still allowing us to solve analytically for the gauge field wave
functions in the background. A shortcoming of this approximation is that $\dot R$ is discontinuous
at the interfaces, which leads to an ultraviolet divergence in the production of gauge bosons.  
The behavior of the actual $R(t)$ is smooth, and must yield a finite amount of particle production
\cite{Cline:Reheating}.  In addition, we will find a separate UV divergence in the particle
production in the limit as the vortex core thickness goes to zero.  This is presumably an artifact
of the effective field theory which is not present in the full string theory, and we deal with it
by  introducing the cutoff $\epsilon$ on the final radius of the defect core, which we will take to
be of order the string length $l_s$.

\subsection{Gauge Field Solutions}

The first step in computing the production of photons in the time-dependent background
is to solve their equation of motion following from (\ref{toyaction}):
\begin{equation}
\label{toyEOM}
  -2{\dot{R}\over R} \dot{A}^{\hat{\rho}} -\ddot{A}^{\hat{\rho}} + {{A^{\hat{\rho}}}''\over R^2} + 
    \frac{{A^{\hat{\rho}}}'}{r R^2} 
   + \frac{\partial_\theta^2 A^{\hat{\rho}}}{r^2 R^2}  + 
   \partial^{\hat{\mu}} \partial_{\hat{\mu}} A^{\hat{\rho}} =0
\end{equation}
where the dot and prime denote differentiation with
respect to $t$ and $r$, respectively.

Equation (\ref{toyEOM}) separates as
\begin{equation}
 A^{\hat{\rho}} (t,r,\theta,x^{\hat{\mu}})=\phi(t) \varphi(r) \Theta(\theta) \chi^{\hat{\rho}}(x^{\hat{\mu}})
\end{equation}
where
\begin{eqnarray}
 \label{largedimensionEOM}
 \partial^{\hat{\mu}} \partial_{\hat{\mu}} \chi^{\hat{\rho}} = 
 -k^2 \chi^{\hat{\rho}},\  
 \partial_{\hat{\nu}} \chi^{\hat{\nu}}=0,\ k^2= k^{\hat{\mu}}k_{\hat{\mu}} 
  = \vec{k} \cdot \vec{k} \\  
  \label{thetaEOM}  \partial_\theta^2 \Theta = -m^2 \Theta           \\
  \label{spaceEOM}    \varphi'' + \frac{1}{r}\varphi' + \left( c^2 - \frac{m^2}{r^2} \right)\varphi =0 \\
  \label{timeEOM} \ddot{\phi} + 2{\dot{R}\over R}\dot{\phi} + \left( {c^2\over R^2} + k^2 \right) \phi =0 
\end{eqnarray}
and $c$ is a separation constant.  

The particular solutions of (\ref{largedimensionEOM}) are labeled by the momenta $\vec{k}$ in the 3 large 
dimensions ($\{x^4, x^5, x^6\}$) and we take them to be normalized according to 
\[
  \int \chi^{\hat{\mu}}_{\vec{k}}  \chi^{\hat{\mu}}_{\vec{k}'} dx^4 dx^5 dx^6 = \delta_{\vec{k} \vec{k}'}.
\]

The particular solution of (\ref{thetaEOM}) is a sum of sines and cosines.  The odd
parity and even parity modes (under $\theta \rightarrow -\theta$) do not mix  with the
even ones,  and for simplicity, we restrict our attention to  the even modes, which
include the massless one.  This can underestimate the effciency of the
reheating by a factor of 2 at most. The solution of
(\ref{thetaEOM}) is thus
\[
  \Theta_m (\theta) = \frac{1}{\sqrt{\pi}} \cos(m \theta).
\]
These are orthogonal for different values of $m$, and requiring that the solution be 
single-valued restricts $m$ to be an integer.

So far we have not been specific about the geometry of the two extra dimensions around
which the original annihilating branes were wrapped.  One simple possibility is a
2-sphere, where the descendant brane and antibrane (vortices) form at antipodal points.
Since all the singular behavior of the tachyon background is localized near these points,
the curvature and topology in the bulk should have little effect on particle production
near the defects.  To simplify the mathematics, we therefore replace either of the 
two hemispheres of the sphere with a flat disk, in the coordinate region $r\le 1$.
The correct boundary condition on radial eigenfunctions of the bulk Laplacian is that
their derivatives vanish at $r=1$, so that they are smooth at the interface where the
two halves of the space are glued together.

The solution of (\ref{spaceEOM}), subject to the boundary condition $\varphi'(r=1)=0$ 
and the requirement that $\varphi(r)$ be regular at the origin, is
\begin{equation}
\label{varphi}
  \varphi_{mn} (r) = \frac{\sqrt{2}}{J_{m+1}(c_{mn})} J_m (c_{mn} r )
\end{equation}
where $c_{mn}$ is the $n$th zero of $J'_m(r)$.  The solutions 
(\ref{varphi}) are orthogonal for different values of $n$.  
The zero mode $n=1$, $m=0$ must be treated separately; it is
the constant solution, where $c_{01}=0$ and $\varphi_{01}(r) = 1$.

The solution of (\ref{timeEOM}) depends on the scale factor.  For $t<0$ and $t>t_c$ the solutions
are trivial and are given by
\begin{equation}
\label{t<0}
  \phi_{mn} (t) = \frac{1}{R_0 \sqrt{2 \omega_{mn}}} \left(  a_{mn} e^{-i \omega_{mn} t}    
                                                       +  a_{mn}^{\dagger} e^{i \omega_{mn} t}   \right)
\end{equation}
and
\begin{equation}
\label{t>t_c}
  \phi_{mn} (t) = \frac{1}{\epsilon \sqrt{2 \bar{\omega}_{mn}}} \left(  d_{mn} e^{-i \bar{\omega}_{mn} t}    
                                                     +  d_{mn}^{\dagger} e^{i \bar{\omega}_{mn} t}   \right)
\end{equation}
respectively.  We have defined  $\omega_{mn}^2 = \frac{c_{mn}^2}{R_0^2} + k^2$ and
$\bar{\omega}_{mn}^2 = \frac{c_{mn}^2}{\epsilon^2} + k^2$. The mutiplicative factors in
(\ref{t<0}) and (\ref{t>t_c}) are introduced for later  convenience and ensure that
$a_{mn}$ and $d_{mn}$ will be properly normalized annihilation operators in the 
appropriate spacetime region when the gauge field is quantized.  It will be
convenient for what follows to introduce phase-shifted annihilation and creation
operators in the region $t>t_c$:  $\bar{d}_{mn} = e^{-i \bar{\omega}_{mn} t_c } d_{mn}$
and  $\bar{d}_{mn}^{\dagger} = e^{i \bar{\omega}_{mn} t_c } d_{mn}^{\dagger}$.  In terms
of these operators (\ref{t>t_c}) becomes
\[
 \phi_{mn} (t) = \frac{1}{\epsilon \sqrt{2 \bar{\omega}_{mn}}} \left(  
                                                  \bar{d}_{mn} e^{-i \bar{\omega}_{mn} (t-t_c)}    
                                  +  \bar{d}_{mn}^{\dagger} e^{i \bar{\omega}_{mn} (t-t_c)}   \right).
\]
We are suppressing the dependence of the annihilation/creation operators
on the 3-momenta $\vec{k}$.

In the region $0<t<t_c$ where the scale factor depends nontrivially on time,
the solution of (\ref{timeEOM}) is
\beqa
\!\!\!\!\!\!\!\!\!\!\!\!\!\!\!\!\!\!\!\!    \phi_{mn} (t) &=& \frac{1}{\sqrt{R_0-\eta t}} 
\left( B_{mn} J_{p_{mn}} \left[ \frac{k}{\eta} (R_0 - \eta t) \right] \right.\nonumber\\
&& \qquad\qquad +   \left. C_{mn} J_{-p_{mn}} \left[ \frac{k}{\eta} (R_0 - \eta t) \right] \right)
\label{0<t<t_c}
\eeqa
where $p_{mn} = \frac{1}{2 \eta} \sqrt{ \eta^2 - 4 c_{mn}^2}$.   Some comments are in
order concerning this solution, which has different behavior for massive and massless
modes.

In the massless case $c_{mn}=0$, $p_{mn} = 1/2$ and $J_p$, $J_{-p}$ are linearly
independent, since $p$ is noninteger.  The constants $B_{mn}$ and $C_{mn}$ in
(\ref{0<t<t_c}) should be interpreted as independent, real-valued constants so that
$\phi_{mn} (t)$ is real.

In the massive case $c_{mn} \not= 0$, $\eta^2 \leq 1$,\footnotemark\  and the order of
the  Bessel functions in (\ref{0<t<t_c}), $p_{mn}$,  is pure imaginary.  Such Bessel
functions are  complex-valued and there are several options for constructing real
solutions.  Since  $(J_{\nu}(x))^\star = J_{\nu^\star} (x)$ for real $x$, we can impose
$C_{mn} = B_{mn}^{\dagger}$.  Calculating the Wronskian of $J_{is}$, $J_{-is}$  verifies
that for real $s\neq 0$ these two solutions are linearly independent.   Since we do not
need to explicitly quantize the field in the region $0<t<t_c$, the dagger can be thought
of simply as complex conjugation.   The interpretation of $B_{mn}$ and $B_{mn}^{\dagger}$
as annihilation and creation operators is unnecessary, since there are no asymptotic
states in this region.

\vspace{3mm}

We summarize this subsection by putting these results together to write the general solution of 
(\ref{toyEOM}) as
\[
  A^{\hat{\rho}} (x^M) = \sum_{m,n} \sum_{\vec{k}}
 \chi^{\hat{\rho}}_{\vec{k}} (x^{\hat{\mu}})  \Theta_m (\theta) \phi_{mn} (t) \varphi_{mn} (r).
\]
where
\begin{widetext}
\[
  \phi_{mn} (t) = \left \{ \begin{array}{ll} 
                 \frac{1}{R_0 \sqrt{2 \omega_{mn}}} \left(  a_{mn} e^{-i \omega_{mn} t}    
                 +  a_{mn}^{\dagger} e^{i \omega_{mn} t}   \right) & \mbox{if $t < 0$}; \\
        \frac{1}{\sqrt{R_0-\eta t}} \left( B_{mn} J_{p_{mn}} [ \frac{k}{\eta} (R_0 - \eta t)] + 
         C_{mn} J_{-p_{mn}} [ \frac{k}{\eta} (R_0 - \eta t)] \right)   & \mbox{if $0 \leq t \leq t_c$}; \\ 
     \frac{1}{\epsilon \sqrt{2 \bar{\omega}_{mn}}} \left(  \bar{d}_{mn} e^{-i \bar{\omega}_{mn} (t-t_c)}    
  +  \bar{d}_{mn}^{\dagger} e^{i \bar{\omega}_{mn} (t-t_c)}   \right)  & \mbox{if $t>t_c$}.  \end{array} 
\right.
\]
\end{widetext}

\subsection{Spectra of Produced Particles}
\footnotetext{One might expect $\eta^2 \leq 1$ on physical grounds.  In fact, for the ensuing arguments to
hold one need only restrict $\eta \leq 3.7$ to ensure that $p_{mn}$ is pure imaginary and nonzero for
all massive modes.}
The next step is to impose continuity of $A^{\hat{\mu}}$ and $\partial_M A^{\hat{\mu}}$ at $t=0$ and $t=t_c$ in  order
to compute the Bogoliubov coefficients, which relate the annihilation and creation operators for $t>t_c$
($\bar{d}_{mn}$ and $\bar{d}_{mn}^{\dagger}$) to those in the region $t<0$ ($a_{mn}$ and
$a_{mn}^{\dagger}$).\footnotemark\ \footnotetext{We are ultimately interested in calculating the number of $d_{mn}$ quanta
in the  vacuum annihilated by $a_{mn}$.  For this purpose it is just as good to use
$\bar{d}_{mn}$,  $\bar{d}_{mn}^{\dagger}$ as $d_{mn}$, $d_{mn}^{\dagger}$ since  the
phase shift cancels out of the number operator.
$d_{mn}^{\dagger}d_{mn}=\bar{d}_{mn}^{\dagger} \bar{d}_{mn}$.} 
\ Smoothness of the solutions at the interfaces is ensured by the continuity of
$\phi_{mn}$ and $\dot{\phi}_{mn}$, since $\phi_{mn}$ is the only part of
$A^{\hat{\rho}}$ which changes between the different spacetime regions.

\subsubsection{Massive Modes}

For the massive modes continuity of $\phi_{mn} (t)$ and $\dot{\phi}_{mn} (t)$ at $t=0$
implies
\begin{equation}
\label{U}
      \left(   \begin{array}{c}
       a_{mn} \\
       a_{mn}^{\dagger} \\
      \end{array} \right) \\
     = 
    U \left(   \begin{array}{c}
        B_{mn} \\
        B_{mn}^{\dagger} \\
      \end{array} \right) \\
    =
    \left[  \begin{array}{cc}
                            u_1 & u_2   \\
                             u_2^{\dagger} & u_1^{\dagger} \\
                                     \end{array} \right]   \\
    \left(   \begin{array}{c}
      B_{mn} \\
      B_{mn}^{\dagger} \\
    \end{array} \right) \\.
\end{equation}
The entries of the matrix $U$ are given by
\begin{eqnarray*}
  u_1 &=& \left( \sqrt{\frac{R_0 \omega_{mn}}{2}} + \frac{i \eta}{2 \sqrt{2 R_0 \omega_{mn}}} \right) 
 J_{p_{mn}} (k R_0/\eta)\\
        &-& i k \sqrt{\frac{R_0}{2 \omega_{mn}}} J'_{p_{mn}} (k R_0/\eta)        \\
  u_2 &=& \left( \sqrt{\frac{R_0 \omega_{mn}}{2}} + \frac{i \eta}{2 \sqrt{2 R_0 \omega_{mn}}} \right) 
 J_{-p_{mn}} (k R_0/\eta)\\
        &-& i k \sqrt{\frac{R_0}{2 \omega_{mn}}} J'_{-p_{mn}} (k R_0/\eta).
\end{eqnarray*}
Continuity at $t=t_c$
similarly gives
\begin{equation}
\label{V}
      \left(   \begin{array}{c}
       \bar{d}_{mn} \\
       \bar{d}_{mn}^{\dagger} \\
      \end{array} \right) \\
     = 
    V \left(   \begin{array}{c}
        B_{mn} \\
        B_{mn}^{\dagger} \\
      \end{array} \right) \\
    =
    \left[  \begin{array}{cc}
                            v_1 & v_2   \\
                             v_2^{\dagger} & v_1^{\dagger} \\
                                     \end{array} \right]   \\
    \left(   \begin{array}{c}
      B_{mn} \\
      B_{mn}^{\dagger} \\
    \end{array} \right) \\.
\end{equation}
The matrix entries $v_i$ are obtained from $u_i$ by replacing $R_0$ with $\epsilon$ and $\omega_{mn}$ with 
$\bar{\omega}_{mn}$.

From (\ref{U}), (\ref{V}) we can write
\[
       \left(   \begin{array}{c}
       \bar{d}_{mn} \\
       \bar{d}_{mn}^{\dagger} \\
      \end{array} \right) \\
     = 
    V U^{-1} \left(   \begin{array}{c}
        a_{mn} \\
        a_{mn}^{\dagger} \\
      \end{array} \right) \\
      = 
        \left[  \begin{array}{cc}
                            \alpha_{mn} & \beta_{mn}   \\
                             \beta_{mn}^{\ast} & \alpha_{mn}^{\ast} \\
                                     \end{array} \right]   \\
    \left(   \begin{array}{c}
      a_{mn} \\
      a_{mn}^{\dagger} \\
    \end{array} \right) \\.
\]
The last equality defines the Bogoliubov coefficients.  Note that there is \emph{no
summation} implied over any of the indices in the above expression.  The indices $m$
and $n$ label the modes of the in- and out-states, 
and the summation which appears in the general definition of the
Bogoliubov coefficients is not present here.

We can now determine the spectrum of Kaluza-Klein (KK) excitations of the photon which
is produced in the tachyon vortex background.  
Observers in the future see a spectrum of massive particles in the final state 
given by
\[
  N_{M>0}^{mn} (k) = | \beta_{mn} |^2 = \frac{1}{|\det U|^2} | u_1 v_2 - u_2 v_1 |^2.
\]
The determinant is
\beqa
  \det U &=& \det V = -\frac{2 i \eta}{\pi} \sin \left( \pi p_{mn} \right) \nonumber\\ &=&
          \frac{2 \eta}{\pi} \sinh \left( \frac{\pi}{2 \eta} \sqrt{4c_{mn}^2 - \eta^2}  \right)
\eeqa
which may be obtained by using the Wronskian of $J_p$ and $J_{-p}$.  The fact that the
two determinants are equal ensures the appropriate normalization $|\alpha_{mn}|^2 -
|\beta_{mn}|^2 =1$ of the Bogoliubov  coefficients. The explicit expression for
$N_{M>0}^{mn} (k)$ can be obtained analytically,  but it is complicated and we do not 
write it out here; instead we will give numerical results.

\begin{figure}
\centerline{\epsfxsize=3.5in \epsfbox{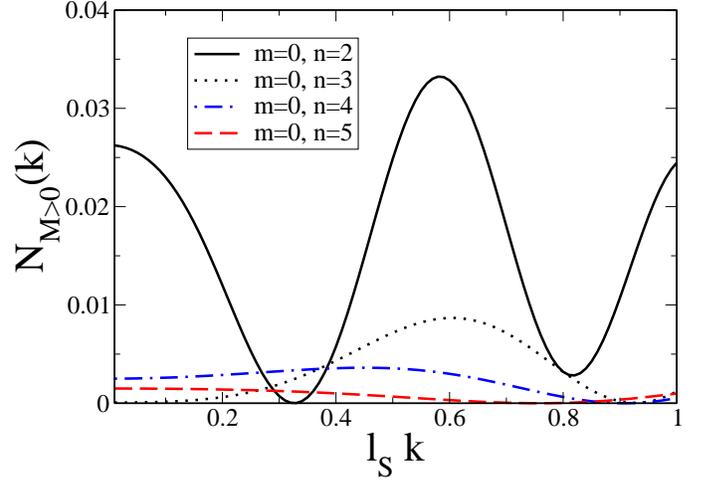}}\caption{{$N_{M>0}^{mn} (k)$ versus $k$ 
for increasing values of the radial quantum number $n$ showing decreased production of heavier modes.}\label{fig3}}
\end{figure}

\begin{figure}
\centerline{\epsfxsize=3.5in \epsfbox{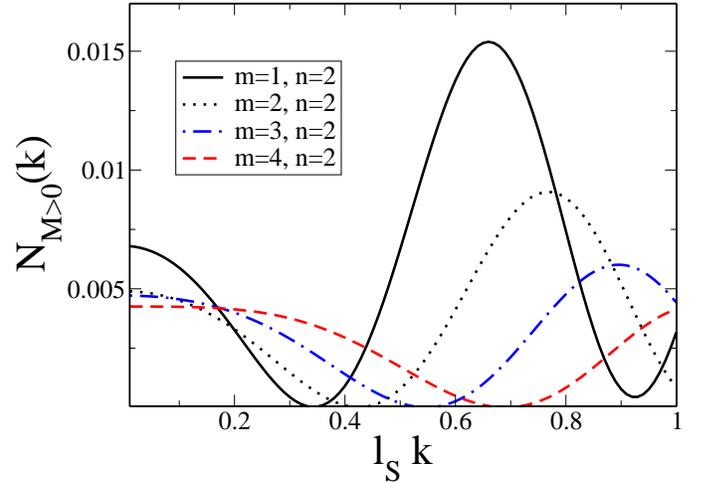}}\caption{
Same as fig.\ \ref{fig3}, but varying the angular quantum number $m$.\label{fig4}}
\end{figure}

The mass of a KK mode with quantum numbers $m,n$ is ${c_{mn}}/{\epsilon}$,  which
increases with $m$ and $n$. Figures \ref{fig3}-\ref{fig4} illustrate the dependence of
$N_{M>0}^{mn} (k)$ on the 3-momentum $k$ for the lightest few massive modes. The
parameters of the vortex background are taken to be $R_0=10 \,l_s$,  $\epsilon = l_s$ and
$\eta=1$.  (The dependence of $N_{M>0}^{mn} (k)$ on $R_0$, $\epsilon$
and $\eta$  is essentially the same as that of the spectrum for massless modes,
$N_{M=0}(k)$, which we will discuss in the next section.)  


\subsubsection{Massless Modes}

\begin{figure}
\centerline{\epsfxsize=3.5in \epsfbox{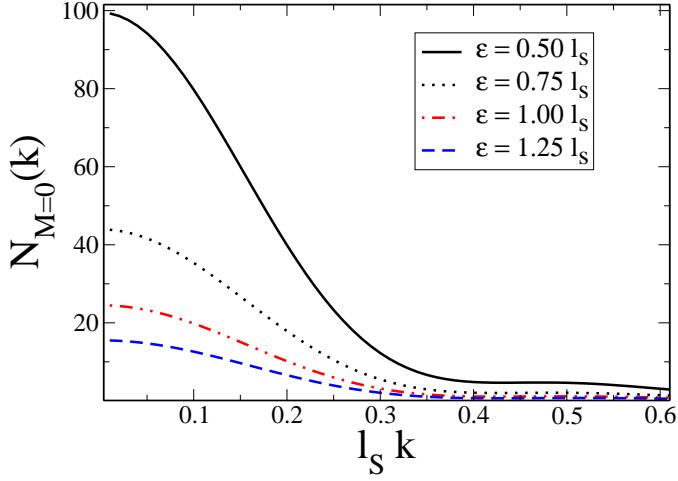}}\caption{{  $N_{M=0} (k)$ versus
 $k$ for different values of $\epsilon$.  $R_0=10 \, l_s$ and $\eta=1$.}\label{fig5}}
\end{figure}

\begin{figure}
\centerline{\epsfxsize=3.5in \epsfbox{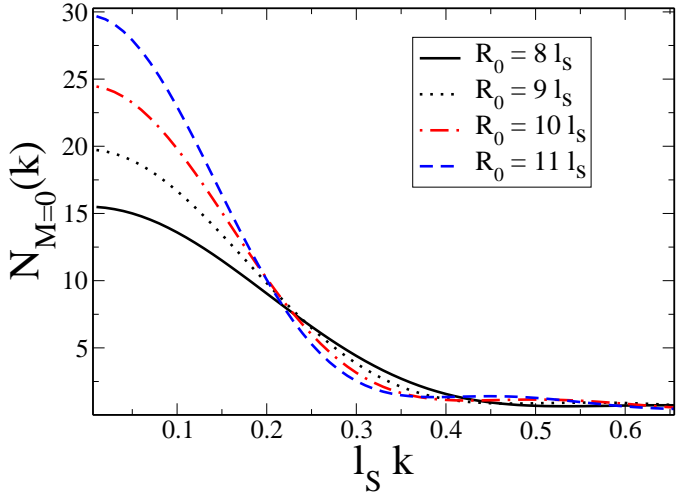}}\caption{{  $N_{M=0} (k)$ versus 
$k$ for different values of $R_0$.  $\epsilon = l_s$ and $\eta=1$.}\label{fig6}}
\end{figure}

\begin{figure}
\centerline{\epsfxsize=3.5in \epsfbox{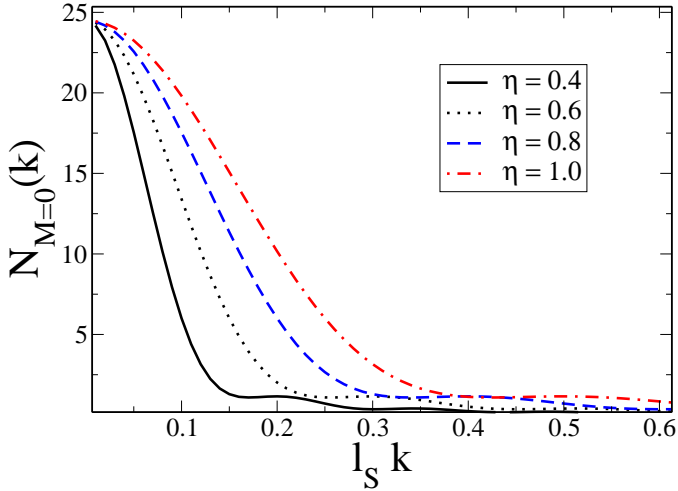}}\caption{{  $N_{M=0} (k)$ versus 
$k$ for different values of $\eta$.  $R_0=10 \, l_s$ and $\epsilon = l_s$.}\label{fig7}}
\end{figure}

The analysis for the massless modes proceeds similarly to the calculations in the preceding
section, but more simply since in this case $p_{mn}=1/2$ and 
$\omega_{mn}  = \bar{\omega}_{mn} = k$.  The Bogoliubov coefficients do not depend on
the mode indices:
\[
       \left(   \begin{array}{c}
       \bar{d}_{mn} \\
       \bar{d}_{mn}^{\dagger} \\
      \end{array} \right) \\
     = 
       \left[  \begin{array}{cc}
                            \alpha & \beta   \\
                             \beta^{\ast} & \alpha^{\ast} \\
                                     \end{array} \right]   \\
    \left(   \begin{array}{c}
      a_{mn} \\
      a_{mn}^{\dagger} \\
    \end{array} \right) \\.
\]
The observed spectrum of particles in the final state
\beqa
\label{nmassless}
  N_{M=0}(k) &=& | \beta |^2 \\
	&\cong&  {1\over \epsilon^2} \, {\eta\pi R_0\over 8 k}\left( J_{\frac12}
	\left({kR_0}/{\eta}\right)^2 + 
	J_{\frac32}(kR_0/\eta)^2 \right) \nonumber
\eeqa
where the second line shows the leading small-$\epsilon$ behavior.  The exact result can also
be obtained in closed form.  Using explicit representations of $J_{n/2}$,
we will be able to integrate this expression exactly to obtain the energy density of produced 
radiation. Figures \ref{fig5}-\ref{fig7} show plots of $N_{M=0} (k)$ as a 
function of $k$ for various values of the parameters $R_0$, $\epsilon$ and $\eta$.

\vskip -3mm

\subsection{Energy Density}

To find the total energy density of radiation produced by the massive modes we should integrate over all
$k$ and sum over the mode indices $m$ and $n$ and the number of polarizations (3 for massive vector
bosons):
\begin{equation}
\label{massiverho}
  \rho_{M>0} = 3\sum_{m,n} \int \frac{d^3 k}{(2 \pi)^3} 
   \sqrt{ k^2 + \frac{c_{mn}^2}{\epsilon^2} }  N^{mn}_{M=0} (k).
\end{equation}
For the massless modes we only integrate over $k$ and sum over the 2 polarizations, since $N_{M>0} (k)$ has no dependence on 
the mode indices:
\begin{equation}
\label{masslessrho}
  \rho_{M=0} = 2\int \frac{d^3 k}{(2 \pi)^3} k N_{M=0} (k).
\end{equation}

As anticipated above, the integrals in (\ref{massiverho}), (\ref{masslessrho}) are not convergent
due to the fact that the simplified tachyon background (\ref{Req}) has discontinuous time
derivatives at $t=0$ and $t=t_c$.   This has the consequence that  the spectrum $N(k)$ decreases
like  $k^{-2}$ for large $k$, which is too slow for convergence.   In reality $n$th derivatives of
$R(t)$ do not exceed $O(1/l_s^n)$, so the production of modes with $k>l_s^{-1}$ should be
exponentially suppressed. We therefore introduce a UV cutoff,  $k_{\rm max} =\Lambda\sim
l_s^{-1}$.  Moreover $N(k)$ is divergent as the vortex radius $\epsilon\to 0$, so $\epsilon$ is
also presumably limited by the string scale, $\epsilon\sim l_s$.   

In the final state, the heavy KK modes have mass given by ${c_{mn}}/{\epsilon}$,  where the
$c_{mn}$'s are order unity and larger.  These are near the cutoff, so their contributions are of
the same order as other UV  contributions which we are omitting. For consistency we should thus
neglect the massive states' contribution to the total energy density on the vortex.  Again, this
underestimates the efficiency of particle production and makes our estimates conservative. 
Henceforth we will refer to $\rho_{M=0}$ as simply $\rho$.

Although the integral (\ref{masslessrho}) can be performed analytically, the  resulting expression
for $\rho$ is cumbersome.  Rather than write it out explicitly we will discuss some
noteworthy features, plot $\rho$ with respect to the  parameters of the model and present
some useful simplifications of the complete expression in various limits.

\begin{figure}
\centerline{\epsfxsize=3.5in \epsfbox{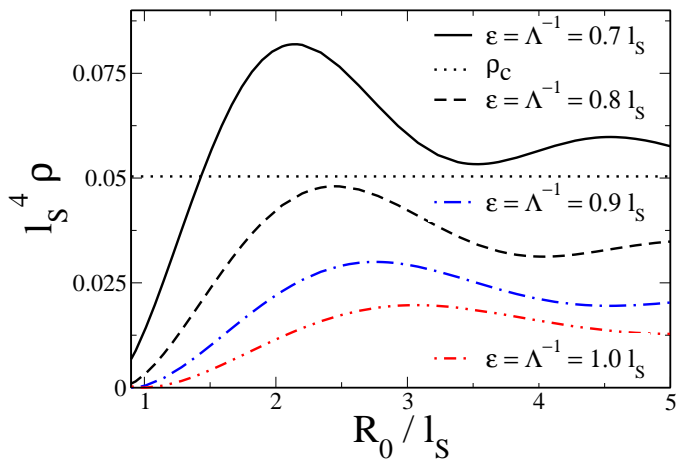}}\caption{{  $\rho$ versus compactification radius
$R_0$ for different
values of $\epsilon=\Lambda^{-1}$ with  $\eta=1$.  Dotted line is the critical density for efficient reheating
$\rho_c$, (\ref{rhoc}). }\label{fig8}}
\end{figure}

\begin{figure}
\centerline{\epsfxsize=3.5in \epsfbox{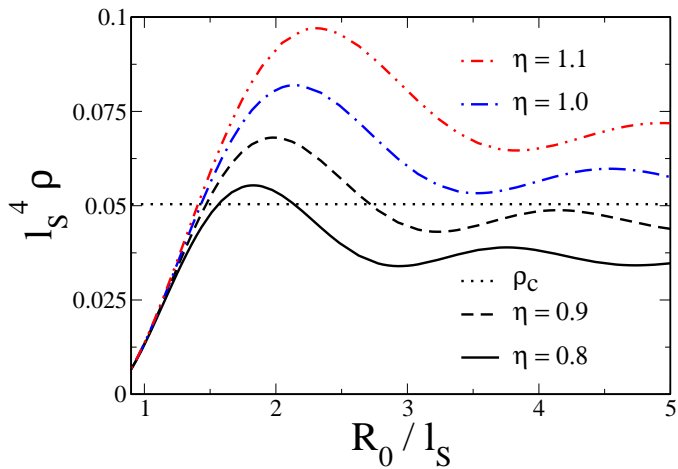}}\caption{{ Same as figure \ref{fig8} but for
different values of vortex collapse rate $\eta$ and with $\epsilon = \Lambda^{-1} = 0.7 \, l_s$.}
\label{fig9}}
\end{figure}

Figures \ref{fig8}-\ref{fig9} show the dependence of $\rho$ on the initial size of the brane.  For
$R_0$ greater than a few times $l_s$, the energy density is relatively insensitive to changes
in $R_0$.  We can therefore reduce the dimensionality of the parameter space by simply assuming
that the extra dimensions are somewhat larger than $l_s$.  In fact in the limit of large $R_0$, the 
energy density takes the very simple form 
\beq
\label{mainresult}
  \lim_{R_0 \to \infty} \rho = \frac{\eta^2}{8 \pi^2} \frac{\Lambda^2}{\epsilon^2}
\eeq
which is the main result of this section.  We recall that $\eta$, which parametrizes the
speed at which the vortex forms, is predicted from eq.\ (\ref{rtilde}) to be $\eta=1$.

As a check on our calculations, we have also considered the limit
as $R_0 \rightarrow \epsilon$, which corresponds to a static background, with no
vortex condensation.  As expected, the energy density of produced particles goes to zero,
\[
  \rho \cong \frac{\Lambda^4}{4 \pi^2
\epsilon^2}(R_0-\epsilon)^2 - \frac{\Lambda^4}{4 \pi^2
\epsilon^3}(R_0-\epsilon)^3  + \dots
\]
as $R_0 \rightarrow \epsilon$.


\subsection{Efficiency of Reheating}

To quantify the efficiency of the reheating we need to determine how much energy is available to
produce the photons on the final-state 3-brane.  Initially the system consisted of a D5-brane plus
antibrane, whose 3D energy density was given by $2\tau_5 V_2$, where $\tau_5$ is the tension of a
D5-brane and $V_2$ is the volume of the compact 2-space $\{ r,\theta \}$ wrapped by the branes. The
final state consists of a D3-brane/antibrane with total tension $2\tau_3$. Until now we considered
just half of the 2-sphere and focused on a single vortex located at $r=0$.   Conservation of
Ramond-Ramond charge requires the second vortex, which we place at the south pole of the sphere to
preserve azimuthal symmetry.   These two defects are identical and are matched at the equator of
the sphere.   The vortex at the south pole represents the D3-antibrane.

Since reheating on each final-state brane should be equally efficient,  the 3D energy density
available for reheating on one them, which we call the critical energy density $\rho_c$,   is half
the difference between the initial and final tensions of the branes and antibranes:
\begin{eqnarray*}
\label{rhoc}
  \rho_c \equiv  \tau_5 V_2 - \tau_3 
              = \tau_3 \left( \frac{V_2}{4 \pi^2 \alphp} -1 \right)
\end{eqnarray*}
where we have used the recursion
relation $2 \pi \sqrt{\alphp} \tau_p = \tau_{p-1}$.  The tension of a D3-brane is given by 
\cite{Polchinski}
\[
  \tau_3 = \frac{1}{g_s} \frac{1}{(2\pi)^3 \alphp^2}.
\]
The string coupling $g_s$, in 5+1 dimensions of which two are compact, is determined
by the gauge coupling evaluated at the string scale, $\alpha (M_s)$ 
\cite{Jones:BraneInteractionsAsOriginOfInflation}:
\[
  g_s = \frac{V_2}{2 \pi^2 \alphp} \alpha (M_s).
\]
Thus we have
\beq
\label{rhoceq}
  l_s^4 \rho_c = \frac{1}{16 \pi^3 \alpha (M_s) } \left(  1 - \frac{4 \pi^2 l_s^2}{V_2}   \right).
\eeq
In the regime where $V_2 \sim 2\pi R_0^2\gg l_s^2$, which was where we could most easily quantify the 
particle production, the second term in parentheses can be neglected, and in any case it
would be unimportant for a rough estimate unless it accidentally canceled the first term ($1$)
to high accuracy.  Hence we drop this term and take $
  l_s^4 \rho_c \cong ({16\pi^3 \alpha (M_s) })^{-1}$.

The energy  density $\rho_c$ is the critical value at which the conversion into radiation would be
100\% efficient. In our analysis we take $\alpha(M_s) \cong \frac{1}{25}$
\cite{Jones:BraneInteractionsAsOriginOfInflation} which gives $\rho_c l_s^4 = 0.05$.  The critical
energy density is shown as a dashed horizontal line in figures \ref{fig8} and \ref{fig9}. We see
that the criterion for efficient reheating can be achieved for moderate values of the parameters. 
We only need for the length-scale cutoffs $1/\Lambda$ and $\epsilon$ to be somewhat  smaller than
the string scale, while the size of the extra dimensions should exceed a few times $l_s$.  Using
(\ref{mainresult}), we can write the criterion for efficient reheating as
\beq
	\sqrt{\epsilon\over\eta\Lambda} \lsim \left({2\pi\alpha(M_s)}\right)^{1/4} l_s  \cong 0.7\, l_s  
\eeq

We have not taken into account the back-reaction of the particle
production on  the tachyon background, which is why our calculation allows for more reheating than
is energetically possible.  The back reaction will suppress somewhat the actual efficiency of
reheating, but we don't expect a dramatic reduction.  Given that we have been conservative in our
estimates, such as ignoring the contributions from  produced KK photons which will decay into
massless photons, our result makes it plausible that a large fraction of the original energy can be
converted into visible radiation.

\section{Summary and Conclusion}

We have argued for the possibility that our visible universe might be a codimension-two brane left
over from annihilation of a D5-brane/antibrane pair at the end of inflation.  In this picture, reheating
is due to production of standard model particles ({\it e.g.} photons) on the final branes, driven by
their couplings to the tachyon field which encodes the instability of the initial state as well
as the vortex which represents the final brane. We find that reheating can be
efficient, in the sense that a sizable fraction of the energy available from the unstable vacuum can
be converted into visible radiation, and not just gravitons.

The efficiency of reheating is greatest  if the radius of compactification of the extra dimensions
is larger than $2$-$3$ times the string length $l_s$.  The efficiency also depends on phenomenological
parameters we had to introduce by hand in  order to cut off ultraviolet divergences in the
calculated particle production rate: namely $\epsilon$, a nonvanishing radius for the final brane,
and $\Lambda$, an explicit cutoff on the momentum of the photons produced.  The latter must be
introduced to correct for discontinous time derivatives in our simplified model of the background
tachyon condensate; the actual behavior of the condensate corresponds to a cutoff of order
$\Lambda\sim 1/l_s$.  It is less obvious why the effective field theory treatment should give
divergent results as the thickness of the final brane ($2\epsilon$) goes to zero, but it seems clear
that a fully string-theoretic computation would give no such divergence, and therefore it is
reasonable to cut off the field-theory divergence at $\epsilon\sim l_s$.  Given only these mild
assumptions, our estimates (\ref{mainresult},\ref{rhoceq}) predict that the fraction of the available energy which is converted
into visible radiation is $\rho/\rho_c \cong \pi\alpha(M_s) \cong 0.25$.  This simple estimate
counts only photons; in a more realistic calculation, it would be enhanced by the number of light
degrees of freedom which couple to the tachyon, which could be much greater 1.  Moreover it could
also be enhanced by the production of massive KK modes, which correspond to string excitations
when the vortex has formed \cite{HN,MZ}.

Our analysis makes significant improvements to the previous work of \cite{Cline:Reheating}.  We
considered the formation of a vortex in the tachyon field rather than a kink, in line with the 
descent relations for stable Dp-branes.  We found analytic solutions for the tachyon field which
give the time dependence in the vicinity of the defect while it is forming, for both the vortex and
the kink solutions.  In the latter case we verified that this solution reproduces the  known
dynamics of kink formation which was determined numerically in \cite{Cline:DbraneCondensation},
giving us more confidence in the vortex solutions, which are quite analogous. Our explicit solution
for the tachyon background is nevertheless too complicated for computing the production of particles
on the defect.  We therefore approximated it by a simpler ansatz with the same qualitative behavior,
which allows for analytic solutions of the gauge fields in  the background.  This ansatz resembles a
gauge field in a 6D spacetime with two compact spatial  dimensions which are contracting with time,
and leads to simple analytic results for the energy density of photons produced during the
contraction, in the regime where the extra dimensions are large compared to the string scale.


There are still some outstanding questions to be addressed concerning this scenario.  First, we have
made reference to the Kibble mechanism for the creation of the final state defects.  If we assume the
causal bound of one defect per Hubble volume then this would imply that the size of the extra dimensions 
must exceed the inverse Hubble rate; otherwise
there would be enough time for the fields to straighten themselves out and the putative
vortex-antivortex pair would immediately annihilate.  For example if we take the string scale $M_s$ to be 
$10^{16}$ GeV then  $H\sim M_s^2/M_p$ and we would need the compactification scale to be of order 
$R_0 \sim (M_p/M_s) l_s$. Our
results indicate that efficient reheating is compatible with a large compactification scale.  However,
taking the remaining four extra dimensions (which the initial 5-brane/antibrane pair do not wrap) to
be string scale is not consistent with getting inflation since the initial state 5-brane and antibrane
cannot be sufficiently separated to satisfy the slow roll conditions.  

One the other hand, our results 
indicate that the reheating \emph{can} be efficient for $R_0$ only a few times the string length, though
in this scenario a naive application of the Kibble mechanism does not favor having the final state defects
span the three large dimensions.  These requirements may not be prohibitive since the question of how these
defects form dynamically at the end of inflation is a quantitative one which merits further investigation.
In principle the correlation length for the initial fluctuations of the tachyon field could be as small as
the string length.  We point out also that it is possible that the dynamics of the formation of tachyon 
defects is qualitatively different from defect formation in a conventional scalar field theory.  
For example, the numerical investigation of
\cite{Cline:DbraneCondensation}, it was found that small kinks in the initial  configuration which
are in causal contact with each other {\it do not} dynamically straighten themselves out as they
would in a conventional, nontachyonic field theory.  Instead, every place where the field crosses
zero in the initial state develops a full-blown kink, so long as there was enough energy in the bulk
to produce the required number of kinks.  Another indication that the dynamics of the tachyon field may be
qualitatively different from an ordinary scalar field theory comes from the \cite{Causal} in which the 
causal structure of the tachyon Dirac-Born-Infeld action was studied.  The authors of \cite{Causal} 
found that small fluctuations of the tachyon field propagate according to an effective metric
which depends on the tachyon background.  In the case of a homogeneous rolling background it was found 
that as the condensation proceeds the effective metric contracts to the Carroll limit of the Lorentz group 
so that the tachyon light cone collapses into a timelike half line and the tachyon fields at different 
spatial points are decoupled.  We feel that quantitatively determining the dynamics of the formation of
tachyon defects at the endpoint of D-brane inflation is a question which deserves further investigation.

\begin{figure}
\centerline{\epsfxsize=3.5in \epsfbox{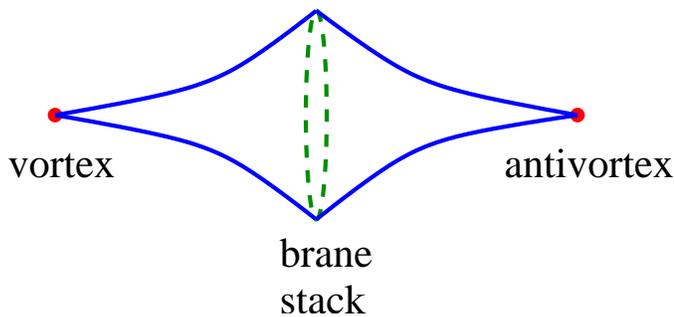}}\caption{Warped compactification with 
branes localized in throats on opposite sides of a stack of branes.
\label{lastfig}}
\end{figure}

In the present analysis we have not included the gravitational or Ramond-Ramond
forces between the vortex and antivortex which would attract them toward each other and lead to
their eventual annihilation.  How do we insure that the braneworld on which we are supposed to live
is safe from annihilation with an antibrane in the bulk?   One possibility is to have warping caused
by a stack of branes which wraps only the equator of the extra dimensions, as illustrated in fig.
\ref{lastfig}.  Such a braneworld scenario using the AdS soliton solution for the bulk has been 
considered in ref.\ \cite{LMW,BCCF,CDGV}.  The advantage for our scenario is that the warping can
provide a barrier to the annihilation of the brane-antibrane pair, since it is energetically
favorable for them to remain within their respective throats.

In solving for the tachyon background we have also ignored the possibility of caustic formation in
the bulk \cite{Felder:Caustics} by taking initial profiles without too much  curvature. It is
possible that caustic formation may be an artifact of the derivative truncation which leads to the 
Born-Infeld type of Lagranian for the tachyon.  See \cite{p-adic} for a discussion of the problems
of dynamical equations with infinitely many derivatives.  

Brane-antibrane inflation and braneworld cosmology are two of the most important applications of
string theoretic ideas to cosmology.  We find it intriguing that these two ideas might be combined
in the way we have described.  An outstanding challenge is to find some observable signatures that
would be able to test our scenario, for example through the gravitational wave component which is
expected to be a major component of the radiation produced during reheating.

We thank Horace Stoica for helpful discussions concerning warped compactifications as well as  Robert 
Brandenberger and Henry Tye for helpful comments regarding the Kibble mechanism.
Our research is
supported by the Natural Sciences and Engineering Research Council (NSERC) of Canada and Fonds
qu\'eb\'ecois de la recherche sur la nature et les technologies (NATEQ).

\end{document}